# The Future of Space Activities and Preservation on Mars: a Preliminary Policy Delphi Study

George Profitiliotis[a,b]*, Jacob Haqq-Misra[a]

[a] Blue Marble Space Institute of Science, Seattle, WA, USA.

[b] Department of Humanities, Social Sciences, and Law, School of Applied Mathematical and Physical Sciences, National Technical University of Athens, Zografou Campus, 15780, Zografou, Athens, Greece.

* Corresponding Author: gprofitil@mail.ntua.gr

Co-author email address: jacob@bmsis.org

**Abstract**

A 'planetary park' system has been suggested as a way for Mars to 'preserve' land for a multitude of purposes. These parks would represent a diverse portion of martian terrain and would be regulated to minimize human contamination and prevent excess human intervention. One suggestion is a two-part land-use policy for Mars, i.e. planetary parks and Lockean land for the use of non-park areas, in which non-park areas could be opened up for development and settlement. Others have proposed that this planetary parks concept could be extended to appeal to corporate interests as a compromise between protecting environmental interests and allowing private use of resources on Mars. This empirical study aims to help advance the core discussion of the interplay between future space activities and future preservation plans on Mars. Specifically, this work uses a particular strategic foresight method, the Policy Delphi, as a means to reveal positions on that issue and to explore alternative policy options in advance, through the collective judgment of a panel of experts. It should be noted that the goal was not strictly to promote consensus regarding the topic of interest, but to encourage structured dialogue. Consequently, the stable consensus achieved in specific areas indicated that the panel of experts foresees international cooperation on matters related to preservation plans for Mars, which, in turn, will enhance the overall international space governance. Furthermore, the panel foresees the inclusion of varying actors, both state and non-state ones, in the development of a preservation framework for Mars, especially on a national level. To this end, the panel endorses the inclusion of a diversity of values to be protected via the aforementioned framework, and suggests the advancement of broad global dialogue on this matter including academic, industrial, governmental, and public actors. The preliminary insights revealed in this work may inform relevant international discussions towards the formulation of proactive policies that will contribute to the environmental governance of future activities on Mars.

**Keywords:** Space Commercialization, Preservation, Mars, Policy Delphi, Futures Studies, Planetary Protection

## 1. Introduction

Access to Mars for both orbital and landed missions appears to be broadening beyond the traditional major space powers of the past few decades. As emerging agencies and new commercial actors are entering the stage, novel issues and considerations come into view regarding the varying activities conducted in space. Indeed, it is foreseeable that the increase and diversification of actors claiming a stake on Mars can lead to contradictory or incompatible interests. That is not without precedent: development activities on Earth, such as industrialization and establishing infrastructure, can cause impacts to the surrounding environment which, in turn, preclude the promotion of other interests. For example, if implemented, visions related to long-term human habitation on Mars and the exploitation of martian resources may irreversibly impact the scientific efforts to search for life on Mars by means of biological contamination of the martian environment with terrestrial biological material. Safeguarding the search for life on Mars by preventing the planet's forward biological contamination has been one of the goals of the planetary protection policy maintained and promulgated by the Committee on Space Research (COSPAR) as a set of nonbinding guidelines for the reference of interested actors (COSPAR, 2021). However, astrobiological scientific knowledge is just one of the goods that could potentially be harmed by the negative impacts of such endeavors, especially in the case of all-private missions to Mars that may benefit from the absence of applicable domestic regulation and may cut down on contamination prevention costs (Profitiliotis & Loizidou, 2019). Interestingly, inspired by the upcoming 'first private mission to Venus' (Rocket Lab, 2022), a newly-formed association has announced a crowdfunding campaign for a privately-funded mission that will investigate the presence of extant indigenous life on Mars before the arrival of the first humans, although the group does 'not believe that we need be too concerned about the risks associated with the planetary protection' (Spacek, 2022).

Nevertheless, Mars is not only valuable for its potential to increase astrobiological scientific knowledge. Recognizing the plurality of values of Mars, Cockell & Horneck (2004) initially suggested a 'planetary park' system for Mars to 'preserve' land for a multitude of purposes, i.e. to protect the intrinsic worth of its geological environment, to safeguard regions of Mars for the enjoyment of future generations, to protect the aesthetic value of sites of natural beauty, to enable the future geological or biological study of representative regions of different terrains, and to protect regions of historical importance because of their relation to past Mars missions. In that sense, 'preservation' of designated planetary park areas would mean their protection from any human alteration, which is stronger than 'conservation', i.e. their protection for intended future use, although some element of a conservationist ethos within such a planetary park system is expected, as these areas might be conserved as a scientific resource for their future use in scientific study (Cockell & Horneck, 2006). These parks would represent a diverse portion of martian terrain and would be regulated to minimize human contamination and prevent excess human intervention (Cockell & Horneck, 2006). Cockell (2006) suggested a two-part land-use policy for Mars, i.e. Planetary Parks and Lockean Land for the use of non-park areas. According to Cockell, non-park areas could be opened up for development and settlement. Bruhns & Haqq-Misra (2016) proposed that this Planetary Parks concept could be extended to appeal to corporate interests as a compromise between protecting environmental interests and allowing private use of resources on Mars.

Given the rapid development of the space sector and the fact that the literature on the most appropriate land-use policy for Mars is still growing (Dapremont, 2021) almost two decades after the introduction of the Mars planetary park system idea, this empirical study aims to help advance the core discussion of the interplay between future space activities and future preservation plans on Mars. Inspired by the observation of Bruhns & Haqq-Misra (2016), who recognized the value of involving the global community of scientists, experts, and leaders in the decision-making process of a potential Mars planetary park system via a consensus-based approach, this work uses the Policy Delphi method as a means to paint a broader picture of the preservation discussion, by revealing opposing positions and by exploring alternative policy options ex ante, through the collective judgment of an interdisciplinary group of experts.

## 2. Methodological background

In an effort to assess the future, RAND researchers created the Delphi method in 1953. The envisioned use of the Delphi method was to augment the practical process of planning by offering enhanced foresight. To this end, the Delphi method generally comprises the formulation of a group of respondents with important expertise in an area of interest and their involvement in multiple repetitive rounds of surveying. As an action-oriented method, the outcomes of its iterative utilization of the participants' expertise are aimed at informing decision-making with insights regarding the future, such as alternative future possibilities, predictions, and probability and desirability of occurrences (Bell, 2009). Studies involving the Delphi method consist of at least eight general steps (Bell, 2009): the specification of the subject whose future will be explored; the creation of a questionnaire to collect data; the construction of a panel of participants with important expertise whose judgments on the subject will be explored; the collection of the participants' initial judgments on the subject via the questionnaire; the aggregation and summarization of the collected data of the initial judgments; the communication of the summarized data of the initial judgments as feedback to all participants; a second round of collecting the participants' subsequent judgments on the subject via the questionnaire, which may have been changed in light of the received feedback; and the analysis, interpretation, and presentation of the data in a final report. It should be noted here that the steps of collecting the participants' judgments, creating a summary of the data, and communicating this summary as feedback to the participants to inform their subsequent judgments may be repeated when there is a need for additional rounds (Bell, 2009).

The Delphi method spread quickly in the academic and practitioner circles, which led to a great number of studies in the literature, both in applications and in methodological innovations that created modified versions of the general method for specific fields of inquiry (Bell, 2009). The Policy Delphi method was originally introduced as one such version of the general Delphi method specifically devoted to examining policy issues for which there are only informed advocates and referees (Turoff, 1970). A major point of departure of the Policy Delphi method from the general Delphi method is that the latter aims to promote consensus among the participants after the multiple iterations. On the contrary, the former aims to generate the strongest possible opposing viewpoints and to uncover and evaluate policy alternatives, which may possibly lead to consensus or dissensus (de Loë, et al., 2016). By not explicitly seeking to promote consensus, the Policy Delphi method can help investigate and analyze policy issues in breadth in order to contribute to decision-making (Meskell, et al., 2014). Another significant difference between the Policy Delphi method and the general one is that in Policy Delphi studies the groups of recruited participants are constructed to be heterogeneous to help reveal as many opinions and viewpoints as possible, while in general Delphi studies the goal is homogeneity which can make it easier to reach consensus (Manley, 2013). Therefore, a Policy Delphi study can be considered a precursor activity to unpack the differing positions advocated with respect to a policy issue and to reveal the main pro and con arguments for those positions. Decision-making committees can then utilize the results of such a study to formulate the required policy (Turoff, 2002).

In general, a Policy Delphi study consists of six phases (Turoff, 2002): the appropriate formulation of the issues to be considered; the exposition of the policy options for the issues under consideration; the determination of the initial positions of the participants on those issues, particularly the ones exhibiting disagreement; the exploration and surfacing of assumptions, views, or facts that are used to support the opposing positions in the case of disagreement; the collective assessment of those underlying rationales that are used to support the opposing positions; and the reevaluation of the options, in light of the revealed rationales that support opposing positions as well as of their collective assessment. Of paramount importance to the appropriate implementation of a Policy Delphi study are the following four principles: anonymity among participants; statistical aggregation of the participants' individual responses; controlled feedback of responses to all participants; and iteration of response collection. These four basic principles enable a structured dialogue to proceed among the participants, leading to the exposition of multiple interpretations and views, while preventing the negative features of group discussions, such as domineering individuals and opinions, which could undermine the whole process (Belton, et al., 2019).

Indeed, the prevention of group think and of the effects of dominant personalities has been reported as one of the major advantages of the Policy Delphi method; other advantages include the avoidance of face-to-face debates that could occur in physical meetings, the provision of adequate time for contemplation on the subject under investigation due to the sequential rounds of data collection and feedback, and the focused and structured analysis of the subject under investigation (Franklin & Hart, 2007).

The Policy Delphi method was selected for the application reported in this article, because of these advantages and characteristics that render it well-suited to assist in the structured exploration of the different positions that exist among experts regarding the interplay between future space activities and future preservation plans on Mars, as well as to enable the ex-ante investigation of alternative corresponding policy options. The design and implementation of the empirical Policy Delphi study that explored the 'Future of Space Activities and Preservation on Mars' is discussed in detail in **Section 3**.

### 3. Design and implementation of the Policy Delphi study

The overall objective of this empirical study was to move forward the discussion of the interplay between future space activities and future preservation plans on Mars in a structured manner, by revealing positions on that issue and by exploring alternative policy options in advance, through the collective judgment of a group of experts. To this end, the Policy Delphi was selected as the most appropriate strategic foresight method for this application, because of its characteristics and its advantages that were presented in **Section 2**. This application closely followed guidelines and recommendations on the design and implementation of Policy Delphi studies that have been published in the academic literature (Turoff, 1970; Turoff, 2002; Bell, 2009; Manley, 2013; Meskell, et al., 2014; de Loë, et al., 2016; Belton, et al., 2019). All the methodological steps in conducting this Policy Delphi study will be presented hereafter, while all the questionnaires and additional documents that were administered to the participants are made available as **supplementary material**. The study plan was reviewed and approved by the Institutional Review Board of the Blue Marble Space Institute of Science (effective date: 15 June 2021).

The process of this study utilized Bell's eight general steps and Turoff's six phases, as presented in **Section 2**, as overall guiding principles. The practical application of this study was based on a six-step methodology that was recently proposed in the literature as a detailed and generalized prescription for a Delphi study workflow, considering, however, the importance of judgment calls to be made by researchers on accommodating those steps to the particular aspects of their specific research cases (Belton, et al., 2019).

The first step was to set up the process by determining the overall goals of the study, by selecting the experts to participate, and by proactively resolving the practicalities of running the study. The overall goals of this study were determined based on the general subjects discussed in **Section 1**, on the authors' open questions, and on the outcomes of an early, exploratory pilot study that was conducted in August 2020 with the voluntary participation of scientists from the Blue Marble Space Institute of Science, which helped to identify the most critical topics related to the future of space activities and preservation on Mars for further investigation. Thus, the following topics were selected for discussion and debate during the exploratory Round 1 of this study: the perceived progress of space activities on Mars and their perceived drivers; the perceived benefits and harms of space activities on Mars; the perceived progress of preservation plans for Mars, their normative dimension, and their perceived value; the perceived options for policy tools to promote preservation plans on Mars and their practical aspects. The Policy Delphi method requires the purposive recruitment of experts based on their expertise and experience related to the topics of interest. In this study, an initial sample frame of 79 experts from academia, governmental agencies, and the private sector was created by the authors, based on their academic or professional work, as reflected in relevant conference presentations, journal articles, books and other published material. It should be noted here that the development of the initial sample frame arrived at the involvement of experts whose affiliated organizations were geographically located in the Western world. To promote complementarity, heterogeneity, and diversity in the academic disciplines represented in the participants' panel that would significantly increase the richness of the rationales and arguments related to the topics of interest, the constructed sample frame included prospective participants with expertise in the following overarching fields: astrobiology and planetary protection; planetary sciences; human exploration and operations; philosophy and ethics; environmental science and conservation; space law and policy; space anthropology and archaeology; space engineering and architecture; space economics and commercialization; and science fiction. Drawing on this sample frame, in total 64 experts were randomly contacted, by means of rolling invitations, until the target of 20 solid and tentative confirmations was reached. It's worth noting that the guidelines of the Policy Delphi method suggest a minimum panel size of 10 (Turoff, 2002) or 7 (Linstone, 1978) experts. Regarding the practicalities of running the study, it was decided to implement the whole process online, considering that the participants were situated in multiple different locations. The study was designed to be conducted in three rounds, each of which would consist of surveying, analyzing responses, and providing feedback to the participants. This number of rounds was deemed sufficient to attain stability in the responses, while avoiding tedious repetition (Bell, 2009) that would demand additional time commitment from the participants thereby increasing drop-out rates (Belton, et al., 2019). The three rounds of this study were to take place over six weeks in July-August 2021, and the whole process was to be conducted in accordance with relevant data protection legislation. Eventually, the practicalities of the implementation went as planned.

The second step was to develop question items and response scales for the questionnaires of the three rounds. Round 1 was exploratory and used 6 open-ended questions to generate concrete issues for further investigation in the subsequent rounds by giving participants the opportunity to express their thoughts on the topics of interest that were discussed in the previous paragraph. A set of 5 demographic questions were also included in the questionnaire. The Round 1 questionnaire is made available as **supplementary material**. Round 2 used 17 statements, thematically grouped into 3 sections, which were formulated to investigate the substantial range of viewpoints on the specific important issues that emerged from the Round 1 responses. Participants were also provided with a condensed aggregate summary of the Round 1 responses as accompanying material. Two kinds of closed-ended, Likert-type response scales were used to capture the participants' judgments on the statements. As Policy Delphi statements are particularly designed to elicit conflict and disagreement and to clarify opinions, response options do not include neutral answers (Turoff, 2002), but can include a 'No Judgment' option, especially to anticipate the case of insufficient expertise for a response (Meskell, et al., 2014). Both response scales allowed for five response options to avoid overwhelming participants with too many options, including a 'No Judgment' option as midpoint, all of which were verbally labeled as follows: 'Strongly Agree', 'Agree', 'No Judgment', 'Disagree', 'Strongly Disagree'; and 'Definitely Feasible', 'Feasible', 'No Judgment', 'Unfeasible', 'Definitely Unfeasible'. In addition, participants were given the opportunity to provide written rationales in open-ended form, as short justifications for their responses. The Round 2 questionnaire and the accompanying summary of the Round 1 responses are made available as **supplementary material**. Round 3 was essentially the iteration of the Round 2 statements, while also incorporating the aggregate ratings provided by the participants in each of the Round 2 statements, as well as a synthesized version of the justifications that were provided for each rating option. Some of the Round 2 statements were reformulated and refined for Round 3, in order to deepen the structured dialogue; these were marked as 'Reformulated Statements'. In addition to the closed-ended, Likert-type response scales, which were the same as in Round 2, participants were given the opportunity to provide open-ended comments or critiques regarding the aggregate ratings, justifications, or content of each statement. The Round 3 questionnaire is made available as **supplementary material**.

The third step was to decide how to deliver the three-round survey to the participants. As the whole process was planned to take place online, it was decided to create all questionnaires on the EUSurvey online platform, which is the European Commission's official survey management tool, and to conduct all communication activities via email. Since there was no automation in this typical Policy Delphi process, the authors manually executed the tasks of releasing the questionnaires, analyzing each round's responses, and providing feedback to the participants between rounds. Participants were asked to provide their email address at the end of each round's questionnaire as a token to facilitate keeping track of the participants' list throughout the rounds of the process; anonymity among participants was maintained in all communications and research stages.

The fourth step was to handle the matters related to providing feedback to the participants between the rounds and to assessing the emergence of consensus. Given the overall objective of this study and the fact that the Policy Delphi method aims not to create consensus but to generate the strongest possible opposing viewpoints on issues within a group of experts and to uncover and evaluate policy alternatives (de Loë, et al., 2016), the feedback was constructed by the authors and was provided to the participants in a structured form to promote critical debate. Round 1 used solely open-ended questions that generated textual responses. Consequently, the constructed feedback of Round 1 towards Round 2 was a condensed aggregate summary of those textual responses. Round 2 used closed-ended, Likert-type response scales that generated nominal responses; it also used requests for open-ended short justifications that generated textual responses. It should be noted here that, because of the fact that, as mentioned above, the Likert-type response scales did not include a neutral option but had 'No Judgment' as a midpoint, the type of categorical data produced by the responses cannot be safely assumed to be ordinal; therefore, it was handled as nominal data. Consequently, the constructed feedback of Round 2 towards Round 3 included for each statement a presentation of the relative frequencies for all the ratings as percentages, as well as a synthesis of the justifications that were provided for those ratings. This feedback was incorporated in the Round 3 questionnaire. It should be noted that although consensus was not an explicit aim of this study, it was still a possible outcome; therefore, the consensus criterion that was established a priori, following (Belton, et al., 2019), was that at least 75% of the participants should provide similar judgment options, e.g. 'Strongly Agree' and 'Agree'. Finally, during the aggregation and synthesis of the textual data, special attention was given by the authors to not disclose any self-reported details by the participants that would compromise their anonymity. **Figure 1** outlines the three rounds of iterative structured dialogue conducted in this Policy Delphi study.

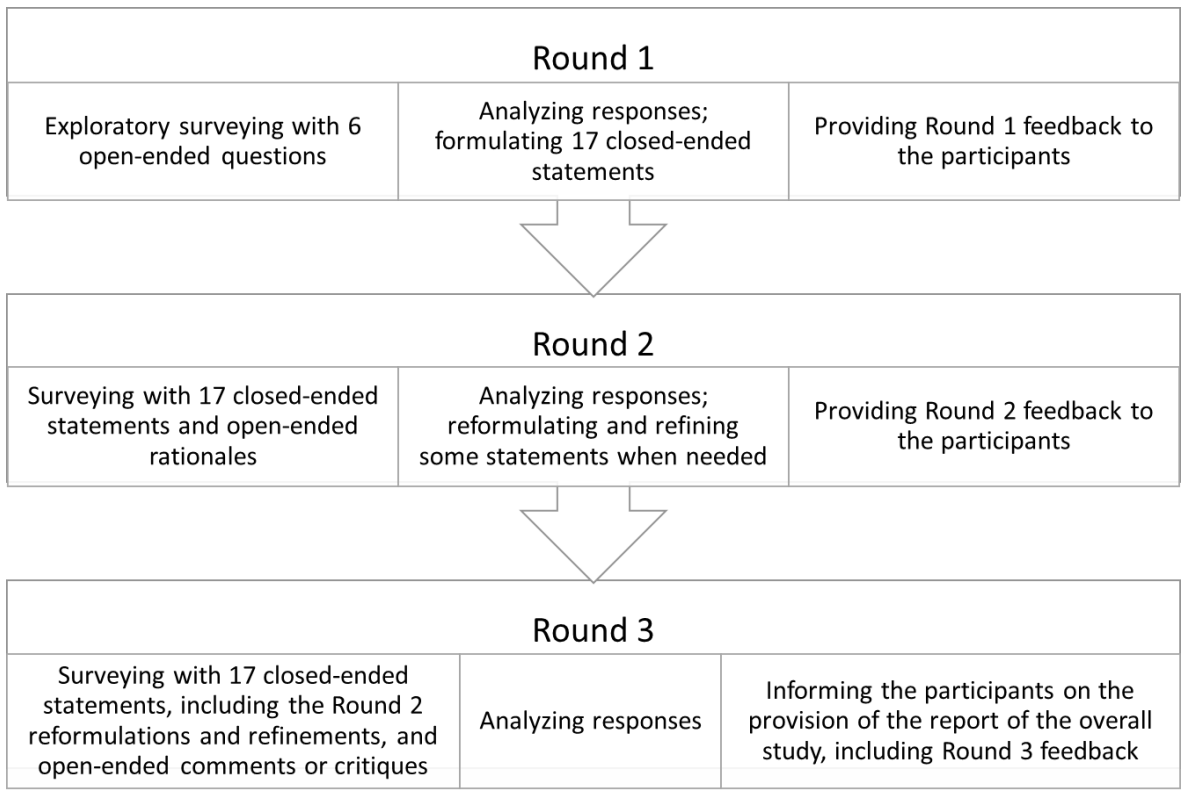

**Figure 1: Outline of the iterative structured dialogue**

The fifth step was to prevent and deal with respondents dropping out of the study. Since participation was voluntary and no reward of any kind was given to the experts, the following measures were taken to prevent attrition. As already mentioned above, the whole process was planned to take place in three rounds over six weeks, while the completion time of each questionnaire was not to exceed 30 minutes, in an effort to keep the necessary time commitment on the participants' side as low as possible without compromising the value of the study. Participants were provided with a clear outline of the commitment required in the original invitation and were reminded of the remaining stages in all subsequent communications. Furthermore, quick turnaround between rounds was used to maintain sustained engagement with the topics of interest; the fact that the participants were purposively selected based on their affinity with these topics also facilitated engagement, potentially because they perceived these topics as important and relevant to them. In addition, communication between and during rounds was conducted manually via emails to maintain a more personal contact with the participants; reminders were sent and short deadline extensions were granted as planned, to accommodate the needs of the participants.

The sixth and final step was to analyze and present the collected data. As mentioned before, respondents were asked to provide responses to both open-ended and closed-ended items in the questionnaires. Therefore, the textual responses to the open-ended items were analyzed using thematic analysis, while the nominal responses to the closed-ended questions were analyzed using descriptive statistics appropriate for this specific type of categorical data. The datasets generated in response to the three questionnaires of this study have been anonymized and are made available as **supplementary material.** During the anonymization, each participant was assigned a unique 'Participant ID' that remains the same across the three datasets. The results of all the analyses of the collected data are presented in **Section 4**.

## 4. Results of the Policy Delphi study

With respect to the response rates, initially 20 invitees confirmed their intent to participate in the study. Afterwards, the registration and consent form was signed by 19 participants; those 19 participants received the Round 1 questionnaire. The Round 1 questionnaire was eventually completed by 17 participants (response rate: 89%) who subsequently received the Round 2 questionnaire. The Round 2 questionnaire was eventually completed by 14 participants (response rate: 82%) who subsequently received the Round 3 questionnaire. The Round 3 questionnaire was eventually completed by 13 participants (response rate: 93%). In each round, the number of participants was above the recommended minimum number of 10 (Turoff, 2002) or 7 (Linstone, 1978) experts. The overall response rate for participants who completed the Round 1 questionnaire and then remained in the process by completing the questionnaires of the other two rounds as well was 76%, which is very typical in the literature (de Loë, et al., 2016). Because of the makeup of the initial sample frame, the participants' affiliated organizations were geographically located in the Western world. It should be noted here that the percentages reported above were rounded to the nearest integer (rounding half away from zero), given the limited panel size. For the same reason, we follow this rounding approach to all percentages reported below as well—due to the rounding, in some cases percentages may not total to 100%.

*4.1 Round 1 results*

The participants' group composition in Round 1 exhibited good heterogeneity. Participants reported being in various stages of their professional careers, ranging from early career roles, to roles of increasing seniority, to retired (mean years of experience in the space sector: 21.88; standard deviation: 13.86). The group's collective professional expertise, as reported by the participants, covered physics, biology, environmental management, evolutionary ecology, astrobiology, planetary protection, innovation, philosophy, space ethics, space law, space policy, social sciences and humanities. In particular, the number of experts who reported professional expertise in each specific field is presented in **Table 1**; it should be noted here that some experts reported more than one fields of professional expertise. Some demographic diversity was also achieved. Specifically, 24% of the participants were 30-39 years old, 29% of them were 50-59 years old, 29% of them were 60-69 years old, and 18% of them were 70 years old and above. Lastly, 71% of the participants reported their sex as 'male', 24% of them as 'female', and 6% as 'other'.

|  | **Number of experts who reported expertise in said fields** | | |
|---|---|---|---|
| **Main fields of professional expertise** | **Round 1** | **Round 2** | **Round 3** |
| Physics | 1 | 1 | 1 |
| Biology | 2 | 2 | 1 |
| Environmental management | 2 | 2 | 2 |
| Evolutionary ecology | 1 | 1 | 1 |
| Astrobiology | 4 | 4 | 4 |
| Planetary protection | 3 | 3 | 3 |
| Innovation | 1 | 1 | 1 |
| Philosophy | 1 | 1 | 1 |
| Space ethics | 4 | 3 | 3 |
| Space law | 1 | 1 | 1 |
| Space policy | 1 | 0 | 0 |
| Archaeology | 1 | 1 | 1 |
| History | 1 | 1 | 1 |
| Languages | 1 | 1 | 1 |
| Social science (in general) | 1 | 0 | 0 |

**Table 1: Panel's disciplinary composition over the three rounds**

As mentioned before, Round 1 explored the following topics for further discussion and debate in the study: the perceived progress of space activities on Mars and their perceived drivers; the perceived benefits and harms of space activities on Mars; the perceived progress of preservation plans for Mars, their normative dimension, and their perceived value; the perceived options for policy tools to promote preservation plans on Mars and their practical aspects. The textual responses that were provided by the participants to the 6 open-ended questions were qualitatively analyzed using thematic analysis (Belton, et al., 2019; de Loë, et al., 2016), in order to identify emergent themes that would guide the formulation of statements for the subsequent rounds. The statements would need to represent different issues that would arise from the textual responses in a way that would promote conservation, but also conflict and disagreement, on the study's subject by highlighting all the obvious questions and sub-issues and encouraging the clarification of opinions (Meskell, et al., 2014). Thematic analysis, i.e. a method of searching across a dataset to identify, analyze, and report recurrent patterned meanings, was conducted inductively under a post-positivist paradigm—i.e. a paradigm which accepts the existence of a reality independent of an observer's thinking that can be studied, but this studying involves the active construction of knowledge by the researcher, which necessarily introduces the inherent limitations of human intellectual mechanisms in the research outcomes—, following a typical six-step framework (Braun & Clarke, 2006; Kiger & Varpio, 2020) that consisted of the following tasks: repeated and active reading of the entire dataset; generating and assigning initial codes, i.e. the most basic elements that can be meaningfully assessed regarding the topic of interest; examining the codes and their underlying data for themes of broader significance that could work together to form a coherent whole; ensuring that codes fit properly their assigned theme in a way that makes themes self-coherent but also distinct from each other, as well as ensuring that individual themes fit meaningfully the entire dataset to represent it; defining and describing each theme, illustrating its relevance to the topic of interest; and reporting the findings.

In this case, textual responses were thematically analyzed, acknowledging that the scope of the Round 1 structured questionnaire was not to provide the material for a self-contained study but to highlight important factors that may affect the expert discussion on the interplay between future space activities and future preservation plans on Mars and to guide the development of contentious statements for the next rounds of the study. The analysis of the responses to all the 6 open-ended questions of the Round 1 questionnaire led to the extraction of some key overarching topics which appear to underlie positions taken in this discourse. These themes are presented and summarized hereafter, while **Table 2** showcases for each theme some selected illustrative examples from the participants' responses verbatim. It should be noted that the early normative and policy-related recommendations of the participants that appeared in this round and were integrated by the authors into their corresponding themes do not constitute consensus views or definitive answers; rather, they are opinions held by participants regarding the sub-issues of the topic of interest which were explicated for in-depth exploration in Round 2 and Round 3.

| Overarching themes | Illustrative examples from the participants' responses |
|---|---|
| Matters of active debate on access to Mars | - 'Space activities on Mars progressing slowly'<br>- 'The progress is impressive' |
| Military outlooks and dominance on Mars | - '[P]ossible appropriation of territory through weaponisation and militarisation of the motion surface or portions of that surface'<br>- 'Mars as a target for political/PR reasons to justify space activities which are really just about escalating militarization/ dominance of space' |
| Advancing Mars science and exploration | - '[N]eed to preserve the opportunity to study the presence or absence of life on Mars'<br>- 'Scientific missions are still the primary driver of Mars missions' |
| Mars as a topic of national and international interest | - '[T]he creation of preservation reserves serves to underline the subversive activity of national appropriation through supposedly legal means'<br>- '[N]eed to ramp up our activities and international dialogues in the realms of ethical and socioeconomic conversations around Mars exploration' |
| Revived visions of settling Mars | - '[R]enewed interest in settling Mars rather than visiting for scientific purposes'<br>- 'One presumed benefit is Mars as an escape pod in case the Earth tips over into catastrophe although I do not agree with this particular perspective' |
| The search for life on Mars | - 'The current planetary protection rules are designed to protect science'<br>- 'If Mars is home to endemic life, then space activities on Mars could harm or even exterminate that life' |
| Human explorers on Mars | - 'Mars exploration will soon be dominated by human missions' |

| | - 'Occasional loss of human life - as with any frontier activity' |
|---|---|
| Prioritization issues of access to Mars | - '[E]nvironmental sustainability of Mars is traded off against other dimensions of benefit'<br>- '[T]he objectives of astrobiology science and potential human missions are not as well synchronized as they might be' |
| Applied science catalyzing and catalyzed by access to Mars | - 'Improving life on Earth'<br>- '[O]n-the-ground research to know which regions are actually Special and which aren't' |
| Inclusivity of values and stakeholders in accessing Mars | - '[S]pace activities on Mars currently tend to reinscribe a colonial, expansionist mindset which isn't all that helpful'<br>- '[T]he many non-scientific, non-instrumental values Mars possesses: intrinsic value; aesthetic value; transformative value; etc.' |
| Societal engagement with access to Mars | - '[R]overs and their way of viewing the landscape and their Twitter feeds etc do strongly encourage people to think that humans belong there too'<br>- '[I]t's crucial that citizens choose to take up the mantle of responsibility for arguing for their own positions in these regards' |
| Economic aspects of access to Mars | - 'Mars is going to be part of any realistic picture for asteroid mining'<br>- '[M]aintain Martian resources through efficient extraction and use' |
| Emerging spacefaring nations accessing Mars | - '[Nations that are not traditional space powers] have activities orbiting Mars'<br>- '[T]here's room for doubt about how stringently [nations that are not traditional space powers] would follow the COSPAR guidelines when their time comes for landers etc.' |
| Private actors targeting Mars | - '[W]e are not yet prepared for the legalities of private industries sending people to Mars before governments do'<br>- '[E]ncourage the private industries involved to join in developing preservation plans' |

**Table 2: Titles of the 14 emerging themes of Round 1 with illustrative examples**

The first theme was 'Matters of active debate on access to Mars'. There is a variety of expert viewpoints regarding the pace of space activities on Mars. A variety of opinions is also evident regarding the magnitude of harm that can be caused by space activities on Mars. Experts also disagree over the need for human Mars missions. There is a diversity of perspectives with respect to the pace and the value of Mars preservation plans. The political factor and the actions of governments regarding Mars preservation are additional points of diverging expert opinions.

The second theme was 'Military outlooks and dominance on Mars'. There is a potential for escalating conflict and exploiting Mars for military purposes. This potential militarization may effectively lead to the appropriation of martian territory. Preservation plans may be used as a benign façade of such appropriation efforts, by becoming a backdoor to exert power.

The third theme was 'Advancing Mars science and exploration'. Planetary science and exploration are major drivers of Mars missions, as they can advance scientific knowledge. However, extensive activities may alter the martian environment in a harmful way, thereby jeopardizing science. Thus, the protection and management of the martian environment that goes beyond the protection of the search for life is important, because it safeguards scientific discoveries. For this reason, the scientific community and scientific non-governmental organizations such as the Committee on Space Research (COSPAR) must be involved in the development of Mars preservation plans.

The fourth theme was 'Mars as a topic of national and international interest'. National and geopolitical motives and international competition are major drivers of Mars missions, although cooperation among nations may also be enhanced through such missions, potentially leading to a decrease in geopolitical conflict. However, Mars surface is also strategically important and nations might try to appropriate it indirectly, by pretending to implement beneficial plans for the protection of its environment. Unfortunately, there is currently insufficient national and international agreement on such plans. On the other hand, this presents the opportunity for Mars preservation to rise as a model of space governance, if successful. For this reason, the United Nations must be involved in the development of Mars preservation plans via binding international agreements, such as a treaty.

The fifth theme was 'Revived visions of settling Mars'. There appears to be a renewed interest for settling Mars in the long term. Nevertheless, there is still disagreement over whether Mars can indeed offer a long-term future for human settlements. In parallel, presenting Mars as a 'back-up' planet may make people less committed to protecting Earth. Settling Mars will require access to resources like water, which is opposed to strict preservation. However, Mars must be protected to be understood as a system, which is necessary for its sustainable future settlement. Indigenous knowledge and experience about long-term living in extreme environments in respectful ways can help develop discussions of settling Mars beyond the visions of a few billionaires that might disregard its protection.

The sixth theme was 'The search for life on Mars'. Searching for life on Mars is an important endeavor that is not pursued sufficiently. This endeavor can help answer existential questions about the origin of life and the place of humanity. However, both the search for life, and perhaps even life itself, may be jeopardized by terrestrial biological contamination. Fortunately, COSPAR's planetary protection policy, implemented by space agencies, is working to protect this endeavor, although it focuses solely on the search. Thus, Mars preservation plans are important for the protection of putative extraterrestrial life, if discovered.

The seventh theme was 'Human explorers on Mars'. Human missions to Mars are envisioned for the future, despite the potential risks to health and even life that may be faced by the crews. For this reason, human missions must be considered in the development of Mars preservation plans.

The eighth theme was 'Prioritization issues of access to Mars'. Gaining scientific knowledge via science and exploration missions to Mars can inform human Mars missions. This creates a prioritization issue of robotic Mars missions over human Mars missions, which, however, extends even further and encompasses robotic missions elsewhere, and other causes that need funds on Earth as well. Although human missions to Mars may yield improved scientific outcomes, their potential for terrestrial biological contamination that can confound science further fuels the tension. There is a matter of trade-off between the protection of martian environment for science, and other interventions and uses, especially human missions.

The ninth theme was 'Applied science catalyzing and catalyzed by access to Mars'. Applied science in the form of research and technology development can lower the barriers for access to Mars. In parallel, new technologies developed in the context of Mars activities can be transferred to terrestrial contexts and solve problems on Earth. Moreover, knowledge gained by studying Mars can be applied in evidence-based policymaking to inform the principles needed for the practical development of plans to protect planetary environments in a consistent manner. Up-to-date scientific understanding is particularly needed in the case of protecting 'Special Regions' on Mars.

The tenth theme was 'Inclusivity of values and stakeholders in accessing Mars'. Current narratives of Mars missions may promote expansionism and colonialism on the part of a few traditional spacefaring nations. However, Mars activities present an opportunity to encourage inclusive cooperation with developing countries for benefit sharing. This inclusion can be extended to indigenous peoples of Earth. Indeed, the scientific value of Mars is just one type of value that has been recognized; other types of non-instrumental values of Mars include ethical, aesthetic, and cultural values. Those values must also be considered as factors for Mars preservation, for a more inclusive approach.

The eleventh theme was 'Societal engagement with access to Mars'. Human curiosity and the need to explore may be important factors for Mars activities. In parallel, Mars may in turn inspire and unite humanity. On the other hand, normalizing Mars activities to the public mind may perpetuate the perspective of Mars as a place to be modified to fit human needs and desires in the long run. The associated pioneer mentality of being first to Mars may be leveraged for various gains. Thus, societal engagement with the issues of Mars preservation is needed for actions to take place, by explicating the reasons for protecting the uniqueness of Mars and taking into account the views of the public. Special consideration must be given to the long-term benefit of future generations; this long-term thinking can also benefit humanity in other matters beyond Mars.

The twelfth theme was 'Economic aspects of access to Mars'. Mars activities can encourage and stimulate economic activities, both in the near-term and in the longer-term future. However, measures to control terrestrial biological contamination are seen as economic obstacles. On the other hand, Mars may possess economically valuable uses and resources that can be exploited, which may be jeopardized by irresponsible access. Thus, preservation plans may help the sustainable management of martian resources.

The thirteenth theme was 'Emerging spacefaring nations accessing Mars'. Nations that are not traditional space powers have commenced Mars activities. These nations may not enforce very stringent regulations for their Mars missions, especially for preservation-related matters.

The fourteenth theme was 'Private actors targeting Mars'. Private actors aspire to play important roles in access to Mars. Their activities are already outpacing the relevant legal frameworks. However, private actors may decide not to follow norms for preservation, if they are not obliged and legally bound. Thus, they must be encouraged to contribute themselves to the development of binding Mars preservation plans.

The results of Round 1 summarized above were used to formulate 17 statements, specifically designed to elicit points of disagreement, which were investigated in more depth during Round 2. **Table 3** showcases how each of the 17 formulated statements was inspired by combinations of the 14 overarching themes.

| Formulated statements for Round 2 | Overarching themes that inspired each statement |
|---|---|
| Statement 1: Totally independent private missions to Mars will be of importance in the future | - Matters of active debate on access to Mars<br>- Revived visions of settling Mars<br>- Societal engagement with access to Mars<br>- Economic aspects of access to Mars<br>- Private actors targeting Mars |
| Statement 2: The progress of space activities on Mars will inflict significant harms in the future | - Matters of active debate on access to Mars<br>- Advancing Mars science and exploration<br>- The search for life on Mars |
|  | - Human explorers on Mars<br>- Prioritization issues of access to Mars<br>- Economic aspects of access to Mars |
| Statement 3: Space activities on Mars will minimize geopolitical conflict | - Matters of active debate on access to Mars<br>- Military outlooks and dominance on Mars<br>- Mars as a topic of national and international interest<br>- Inclusivity of values and stakeholders in accessing Mars<br>- Emerging spacefaring nations accessing Mars |
| Statement 4: Preservation plans for Mars will enhance international space governance | - Matters of active debate on access to Mars<br>- Military outlooks and dominance on Mars<br>- Advancing Mars science and exploration<br>- Mars as a topic of national and international interest<br>- Inclusivity of values and stakeholders in accessing Mars<br>- Emerging spacefaring nations accessing Mars |
| Statement 5: Space activities on Mars with the goal of actually searching for life should progress faster | - Matters of active debate on access to Mars<br>- Advancing Mars science and exploration<br>- The search for life on Mars<br>- Prioritization issues of access to Mars |
| Statement 6: Research and technology development activities to enable crewed missions to Mars should progress faster | - Matters of active debate on access to Mars<br>- Advancing Mars science and exploration<br>- Human explorers on Mars<br>- Prioritization issues of access to Mars<br>- Applied science catalyzing and catalyzed by access to Mars |
| Statement 7: National governments should have stronger political motivation and will for space activities on Mars | - Matters of active debate on access to Mars<br>- Military outlooks and dominance on Mars<br>- Mars as a topic of national and international interest<br>- Emerging spacefaring nations accessing Mars |
| Statement 8: The involvement of the private industry in space activities on Mars should be increased | - Matters of active debate on access to Mars<br>- Prioritization issues of access to Mars<br>- Economic aspects of access to Mars<br>- Private actors targeting Mars |
| Statement 9: Humanity should establish long-term presence on Mars to become multiplanetary | - Matters of active debate on access to Mars<br>- Revived visions of settling Mars<br>- Human explorers on Mars<br>- Prioritization issues of access to Mars<br>- Societal engagement with access to Mars |
| Statement 10: Preservation plans for Mars should include more values than just the protection of astrobiological science from biological contamination | - Matters of active debate on access to Mars<br>- Advancing Mars science and exploration<br>- The search for life on Mars<br>- Prioritization issues of access to Mars<br>- Inclusivity of values and stakeholders in accessing Mars<br>- Societal engagement with access to Mars |
| Statement 11: Broad and global dialogue among governments, private industry, and diverse public | - Matters of active debate on access to Mars<br>- Mars as a topic of national and international interest |
| groups regarding preservation plans for Mars should be initiated promptly | - Inclusivity of values and stakeholders in accessing Mars<br>- Societal engagement with access to Mars<br>- Emerging spacefaring nations accessing Mars<br>- Private actors targeting Mars |
| Statement 12: International governmental actors reach collective agreement on how to conduct Mars activities within a preservation framework for Mars | - Matters of active debate on access to Mars<br>- Military outlooks and dominance on Mars<br>- Mars as a topic of national and international interest<br>- Inclusivity of values and stakeholders in accessing Mars<br>- Emerging spacefaring nations accessing Mars |
| Statement 13: National governments work on national space legislation regarding a preservation framework for Mars | - Matters of active debate on access to Mars<br>- Mars as a topic of national and international interest<br>- Private actors targeting Mars |
| Statement 14: Non-state actors, including commercial actors and NGOs, become involved in the development of a preservation framework for Mars | - Matters of active debate on access to Mars<br>- Societal engagement with access to Mars<br>- Private actors targeting Mars |
| Statement 15: Public support for a preservation framework for Mars is earned through a broad consultation on a global level with experts in Mars activities and non-experts, including indigenous people | - Matters of active debate on access to Mars<br>- Mars as a topic of national and international interest<br>- Inclusivity of values and stakeholders in accessing Mars<br>- Societal engagement with access to Mars |
| Statement 16: COSPAR's Panel on Exploration, advised by the Panel on Planetary Protection, undertakes oversight of a preservation framework for Mars | - Matters of active debate on access to Mars<br>- Advancing Mars science and exploration<br>- The search for life on Mars<br>- Prioritization issues of access to Mars<br>- Applied science catalyzing and catalyzed by access to Mars |
| Statement 17: An Interplanetary Space Authority is created as a specialized UN agency responsible for governing a preservation framework for Mars | - Matters of active debate on access to Mars<br>- Military outlooks and dominance on Mars<br>- Mars as a topic of national and international interest<br>- Prioritization issues of access to Mars<br>- Emerging spacefaring nations accessing Mars |

**Table 3: Mapping of the 17 formulated statements to the overarching themes**

*4.2 Round 2 results*

Despite the attrition from Round 1, the participants' group composition in Round 2 also exhibited good heterogeneity. The attrition had little effect on the representation of various career stages (mean years of experience in the space sector: 24; standard deviation: 14.23). The synthesis of the group's collective professional expertise, as previously reported by the participants, was minimally affected thanks to existing overlapping, as presented in **Table 1**. Demographic diversity was also slightly affected: 21% of the participants were 30-39 years old, 21% of them were 50-59 years old, 36% of them were 60-69 years old, and 21% of them were 70 years old and above. Lastly, 71% of the participants had reported their sex as 'male', 21% of them as 'female', and 7% as 'other'.

The 17 statements that were judged by the participants were grouped into three sections: 'expectations' (statements 1 to 4), 'preferences' (statements 5 to 11), and 'policy perspectives' (statements 12 to 17). As the responses to the 17 statements generated nominal categorical data, their descriptive statistics presented hereafter are limited to their relative frequencies. Moreover, the open-ended form rationales that were provided as short justifications for the responses generated textual data that was analyzed using the initial coding steps of thematic analysis; the distilled coded arguments extracted from this data are presented hereafter for each statement.

Regarding 'Statement 1: Totally independent private missions to Mars will be of importance in the future', the main findings are as follows. The majority position was 'Agree' (36%), followed by 'Strongly Agree' (29%); a 'No Judgment' position was expressed by 21% of the respondents, while 7% of them stated that they 'Disagree', and 7% of them stated that they 'Strongly Disagree' with this statement.

The arguments justifying agreement with this statement were that:

- private missions may catalyze innovation by testing technologies and survival strategies;
- there is already a rising number of motivated established space companies and emerging space startups with rapid and successful track-records;
- private activities will require governmental oversight and regulation;
- governments will still be required to sponsor technology development by the private industry.

The arguments justifying disagreement with this statement were that:

- it will take decades for such missions to happen and to become 'important' in either the financial or the exploration sense;
- it would be unacceptable to allow totally independent, unsupervised private movement of populations in space, because history has shown that such cases are connected with exclusive, oppressive, and undemocratic practices.

Refraining from judgment was justified by the following views:

- such private activities are conceivable but unlikely any time soon;
- such activities may be 'important' in the sense of 'impactful' businesswise, but not 'important' in the sense of 'valuable' for humanity's future in space, as they may harmfully interfere with scientific exploration.

Regarding 'Statement 2: The progress of space activities on Mars will inflict significant harms in the future', the main findings are as follows. The majority position was 'No Judgment' (50%), followed by 'Agree' (29%); 14% of the respondents stated that they 'Strongly Disagree', and 7% of them stated that they 'Disagree' with this statement.

The arguments justifying agreement with this statement were that:

- humans may cause significant pollution, damage, modifications, and other detriments to the martian environment, including any potential extant life;
- the harsh environment of Mars, including potential extant life, may negatively impact missions;
- a violation or loosening of planetary protection requirements in the implementation of Mars missions can exacerbate potential threats.

The arguments justifying disagreement with this statement were that:

- it depends on how 'significant harms' are defined (e.g. to humans, or to the martian environment, or to international activities);
- current and upcoming measures for responsible exploration will catch up and effectively prevent harms;
- there is probably no extant life to be harmed, the martian environment is immune to any reasonable human activity, and the loss of few astronauts' lives is not 'significant harm'.

Refraining from judgment was justified by the following views:

- if proactively managed by means of oversight and regulations, studying Mars and even human presence on Mars can be implemented without significant, or with reduced, environmental harm;
- it depends on the definition of 'significant harms', because there will be impacts but their significance and harmfulness is uncertain.

Regarding 'Statement 3: Space activities on Mars will minimize geopolitical conflict', the main findings are as follows. The majority position was 'Disagree' (50%), followed by 'Strongly Disagree' (21%); a 'No Judgment' position was expressed by 14% of the respondents, while 14% of them stated that they 'Agree' with this statement.

The arguments justifying agreement with this statement were that:

- there is a tendency for shared progress in space to reduce or mitigate conflict between the sharing partners;
- space exploration and long-term settlement will enable understanding and dialogue across nations, on condition that a variety of nations partake equitably in such plans.

The arguments justifying disagreement with this statement were that:

- past experience in shared space activities shows that geopolitical conflict will be reduced but not minimized;
- space activities play only a small role in terrestrial geopolitics, so space activities until now have not reduced geopolitical conflict and neither will Mars activities;
- Mars provides another platform for competition for limited resources, so, judging by our experience with the high seas, geopolitical conflicts will be increased;
- history has shown that conflicts are exported by governments to new frontiers, so geopolitical conflict will be maximized and extended into space, as demonstrated by the establishment of many space forces.

Refraining from judgment was justified by the following views:

- if advances in other sectors are useful analogies, there will always be some geopolitical conflict, but also cooperation;
- in many ways Mars activities will minimize geopolitical conflict, but they could also increase it in the case of a new space race.

Regarding 'Statement 4: Preservation plans for Mars will enhance international space governance', the main findings are as follows. The majority position was 'Agree' (50%), followed by 'Strongly Agree' (29%); a 'No Judgment' position was expressed by 7% of the respondents, while 7% of them stated that they 'Disagree', and 7% of them stated that they 'Strongly Disagree' with this statement.

The arguments justifying agreement with this statement were that:

- Mars preservation plans of any form provide a readily understood rationale for co-operation, so the way that Mars exploration will be conducted will become a precedence and an example for overall international space governance frameworks;
- potential viable preservation plans will be the result of enhanced international space governance rather than its cause, but preserving Mars would in turn enable keeping humanity's options open for its future;
- proactive planning for the long-term exploration and settlement of Mars will improve anticipatory planning in a variety of realms on the part of nations.

The arguments justifying disagreement with this statement were that:

- the creation of sites for preservation on Mars will allow access for only a very small approved minority of actors, which will be undemocratic and unfair;
- the agreement on Mars preservation plans in the first place will only interest a minority fraction of countries on Earth.

Refraining from judgment was justified by the following view:

- there is a marked effort by space agencies to keep international space governance from limiting private plans for future Mars missions.

Regarding 'Statement 5: Space activities on Mars with the goal of actually searching for life should progress faster', the main findings are as follows. The majority position was 'Strongly Agree' (43%); a 'No Judgment' position was expressed by 21% of the respondents, while 21% of them stated that they 'Disagree', and 14% of them stated that they 'Agree' with this statement.

The arguments justifying agreement with this statement were that:

- there is great wealth of scientific knowledge and understanding that can be gained regarding the origin and extent of life in the universe by exploring Mars, which may be both instrumentally and intrinsically valuable, especially if it represents a second genesis;
- the means to accomplish this exist, and there are still many instruments to use for the first time;
- human presence on Mars will quickly disperse terrestrial biological contamination in the atmosphere and on the surface of the planet, so searches for past or present life on Mars should be enhanced before this occurs;
- the gradual 'normalization' of the space sector may lead to it becoming less open towards including diverse activities

The arguments justifying disagreement with this statement were that:

- there is no overwhelming need to find extraterrestrial life;
- the rate should be slower and cautious to allow for new scientific insights to be derived and to be considered for future missions, as there will be no going back from contaminating potential life on Mars;
- Mars sample return to Earth poses a great question.

Refraining from judgment was justified by the following views:

- it is hard to rationalize the search for life on Mars;
- most of humanity has more immediate needs that are at least as important as the search for life on Mars.

Regarding 'Statement 6: Research and technology development activities to enable crewed missions to Mars should progress faster', the main findings are as follows. The majority position was 'No Judgment' (36%), followed by 'Strongly Agree' (29%); 21% of the respondents stated that they 'Disagree', and 14% of them stated that they 'Agree' with this statement.

The arguments justifying agreement with this statement were that:

- the means to accomplish human Mars missions exist, only funding is needed;
- there are still many knowledge gaps to be investigated in the existing R&D roadmaps, especially with respect to surviving, staying safe, and being able to perform effectively on Mars;
- crewed missions are needed to study polar regions of Mars as analogues for the benefit of polar science on Earth;

- crewed missions are needed, because the discovery of extant life that represents a second genesis will change everything.

The arguments justifying disagreement with this statement were that:

- there is no overwhelming need to enable crewed missions to Mars and it is not a priority;
- we are far from the point of diminishing scientific returns from uncrewed in situ and remote studies of Mars;
- we do not need a Hollywood type 'show' mission.

Refraining from judgment was justified by the following views:

- there are reasons to both go faster while we can and to go slower as we don't have enough knowledge regarding such activities, especially their effects on the environment of Mars;
- the pace should be set by the governments who should also determine the affordability of Mars exploration;
- most of humanity has more immediate needs that are at least as important as R&D for crewed missions.

Regarding 'Statement 7: National governments should have stronger political motivation and will for space activities on Mars', the main findings are as follows. The majority position was 'Disagree' (36%); a 'No Judgment' position was expressed by 21% of the respondents; 21% of them stated that they 'Agree', while 14% of them stated that they 'Strongly Agree', and 7% of them stated that they 'Strongly Disagree' with this statement.

The arguments justifying agreement with this statement were that:

- national governments should support scientific exploration activities on Mars, but not settlement or resource exploitation;
- history has shown that political motivation can surely accelerate space activities;
- Mars activities are part of a broader series of technological changes which may well have a civilization-level impact and may encourage society to adopt long-term thinking.

The arguments justifying disagreement with this statement were that:

- there is no overwhelming need for Mars activities, especially given the need to take care of Earth which is the only planet that can sustain humans;
- Mars will not be profitable anytime for national governments and will not offer them further capability;
- political motivation and will may be nationalistic or military in nature, which will be unacceptable as a basis for Mars activities.

Refraining from judgment was justified by the following views:

- this is an issue of budget prioritization to be decided by national governments;
- this statement is too vague.

Regarding 'Statement 8: The involvement of the private industry in space activities on Mars should be increased', the main findings are as follows. The majority position was 'Agree' (36%), followed by 'Disagree' (29%); a 'No Judgment' position was expressed by 21% of the respondents, while 14% of them stated that they 'Strongly Agree' with this statement.

The arguments justifying agreement with this statement were that:

- private investment portfolios can include difficult Mars missions;

- Mars exploration and tourism will motivate the increased involvement of the private industry;
- more private sector involvement means better chances for more significant progress and innovative contributions by competition and/or collaborations;
- private industry involvement also means involvement in preservation plans.

The arguments justifying disagreement with this statement were that:

- the current arrangement of private industry servicing national space programs through competitive bids works well at the moment and should be like that;
- increased involvement of private industries in Mars activities can have negative consequences in promoting late capitalist values as the only way to progress in space;
- the private space industry is restricted to a tiny minority of extremely wealthy persons who are supported by both government and private funds, which makes this involvement inequitable and undemocratic, particularly in the case of their potential involvement in selecting scientific objectives.

Refraining from judgment was justified by the following views:

- it is difficult to see either ideological or commercial motives for such an involvement, besides motives provided by a national government;
- this statement is too vague.

Regarding 'Statement 9: Humanity should establish long-term presence on Mars to become multiplanetary', the main findings are as follows. The majority position was 'Strongly Agree' (29%); a 'No Judgment' position was expressed by 21% of the respondents; 14% of them stated that they 'Agree', while 14% of them stated that they 'Disagree', and 14% of them stated that they 'Strongly Disagree' with this statement. It's worth noting that 7% of the respondents refrained from stating an answer to this statement.

The arguments justifying agreement with this statement were that:

- Mars is one destination in space that can be seen as a rational defense to 'long-term' threats to the Earth;
- the Moon is a good testing ground, but Mars has more space;
- humans are explorers and wanderers, and these attributes will be applied to Mars as well;
- migration is part of the normal activity of humans and can be extended into space with the goal of establishing a multiplanetary presence.

The arguments justifying disagreement with this statement were that:

- although humans should belong to a larger region of space than the Earth, the multiplanetary idea and framing is a little misleading;
- establishing colonies and living out human life on Mars is neither possible not necessary anytime soon;
- only the long-term presence of scientists could bring benefits.

Refraining from judgment was justified by the following views:

- establishing long-term presence on Mars is irrelevant in the nearest future;
- it may not be possible for physical and biological reasons that need to be scientifically determined;
- disagreement with this utilization of 'should'.

Regarding 'Statement 10: Preservation plans for Mars should include more values than just the protection of astrobiological science from biological contamination', the main findings are as follows. The majority position was 'Strongly Agree' (64%), followed by 'Agree' (29%); a 'No Judgment' position was expressed by 7% of the respondents.

The arguments justifying agreement with this statement were that:

- planetary protection is insufficient should there be competing national or commercial interests at Mars;
- planetary protection is time-limited by the 'period of biological exploration', while preservation plans are timeless;
- non-astrobiological science and environmental/resource stewardship depend on other Mars characteristics and dimensions that need to be protected other than potential biological ones, such as planetary data sources (including rare geologic evidence), planetary resources (including water), significant physical features;
- preservation includes more values than pure science (safeguarding historical, aesthetic, and other not just scientific planetary assets);
- current understanding of what should be protected stems from various 20th century discourses which offer only a limited insight into the range of things that humans are likely to value in the future;
- Mars should be seen not just as a resource for humans but as an entity in a solar system setting that has values in its own right.

Refraining from judgment was justified by the following view:

- it is difficult to balance political and economic ideologies and approaches with cultural and scientific values.

Regarding 'Statement 11: Broad and global dialogue among governments, private industry, and diverse public groups regarding preservation plans for Mars should be initiated promptly', the main findings are as follows. The majority position was 'Agree' (50%), followed by 'Strongly Agree' (43%); 7% of the respondents stated that they 'Disagree' with this statement.

The arguments justifying agreement with this statement were that:

- before any plans can be developed, international consensus must be sought for the need for preservation;
- as the possibility of such exploration and the need for dialogue arises, discussions should occur;
- preservation plans must be initiated prior to potential adverse impacts to Mars and before actors are invested, as they may be more flexible and prone to agree to such measures at early stages;
- discussions for long-term strategy must begin now, to avoid burdening future generations with the problems that would arise otherwise;
- diverse stakeholders for effective dialogue should include more than just the scientific communities, particularly the majority of populations and cultures of the Earth that are not usually represented.

The argument justifying disagreement with this statement was that:

- although not problematic, such discussions should not be mandated, as voluntary industry consensus standards seem to be most appropriate.

Regarding 'Statement 12: International governmental actors reach collective agreement on how to conduct Mars activities within a preservation framework for Mars', the main findings are as follows. The majority judgment was 'Feasible' (64%), followed by 'Definitely Feasible' (29%); 7% of the respondents found this statement 'Definitely Unfeasible'.

The arguments justifying the feasibility of this statement were that:

- this has been demonstrated in the past through other activities like the Antarctic Treaty, Aviation Law, the Law of the Sea, COSPAR, etc.;
- the development of a workable framework by governmental actors is possible, but it is being prevented by political will and a misplaced fear of benefit sharing;
- although a single agreement might appear unfeasible in the present geopolitical climate, given China's emerging hegemony in space, there are long-term pressures towards a collective agreement beyond just loose sets of norms and localized instruments;
- it is realistic to reach collective agreement on minimizing damage, although preservation per se is unlikely;
- long-term thinking was common among some of our ancestors and can be promoted on national and international scales, if there is will;
- dominant actors are likely to dominate the different political and economic ideologies and cultural values of the many nations on Earth, as shown in the Artemis Accords.

The argument justifying the unfeasibility of this statement was that:

- 'preservation' is a loaded term which, for many people, means no change and no use, usually facing considerable resistance, so 'conservation' is more appropriate and would be feasible.

Regarding 'Statement 13: National governments work on national space legislation regarding a preservation framework for Mars', the main findings are as follows. The majority judgment was 'Feasible' (43%), followed by 'Definitely Feasible' (21%); a 'No Judgment' position was expressed by 14% of the respondents, while 14% of them found this statement 'Unfeasible'. It's worth noting that 7% of the respondents refrained from stating an answer to this statement.

The arguments justifying the feasibility of this statement were that:

- each nation would want its own perspective included in any internationally agreed framework, so this could be a precursor to a collective agreement;
- the associated problems are just the usual ones of national-level legislation;
- major space powers can lead by the example of their national frameworks;
- this kind of legislation could be applied to all government, public and private actors planning Mars missions, strengthening the application of the Outer Space Treaty to Mars;
- thought leaders in the space exploration industry can commence this work by garnering interest among the public in order to voice their interests to their politicians.

The arguments justifying the unfeasibility of this statement were that:

- 'preservation' is a loaded term which, for many people, means no change and no use, usually facing considerable resistance, so 'conservation' is more appropriate and would be feasible;
- it will not be feasible for a long time, and it should be governments and national agencies thinking about international issues and concerns.

Refraining from judgment was justified by the following view:

- this will be feasible for some countries, but others will resist a framework that could be perceived as restrictive in any way.

Regarding 'Statement 14: Non-state actors, including commercial actors and NGOs, become involved in the development of a preservation framework for Mars', the main findings are as follows. The majority judgment was 'Definitely Feasible' (50%); a 'No Judgment' position was expressed by 21% of the respondents, while 21% of them found this statement 'Feasible', and 7% of them found it 'Definitely Unfeasible'.

The arguments justifying the feasibility of this statement were that:

- it is in their best interests to become involved, but they should not be the sole responsible for defining preservation plans and policing them, as their prime responsibility is towards their shareholders;
- this dialogue among all stakeholders is far more likely in the short term and will be a better driver for effective international collaboration, but consensus must first be sought for the need for preservation;
- it is important to have the input of NGOs who play an important role in setting standards and norms of behavior, and private actors should also be encouraged to have buy-in to a preservation framework, to demonstrate that they are good space citizens;
- building this in from the beginning and emphasizing that preservation is part of sustainability is crucial;
- the Antarctic Treaty and tourists in Antarctic show how this can be done.

The argument justifying the unfeasibility of this statement was that:

- 'preservation' is a loaded term which, for many people, means no change and no use, usually facing considerable resistance, so 'conservation' is more appropriate and would be feasible.

Refraining from judgment was justified by:

- insufficient knowledge of this matter on the respondents' part.

Regarding 'Statement 15: Public support for a preservation framework for Mars is earned through a broad consultation on a global level with experts in Mars activities and non-experts, including indigenous people', the main findings are as follows. The majority judgment was 'Definitely Feasible' (36%); a 'No Judgment' position was expressed by 29% of the respondents, while 21% of them found this statement 'Feasible', and 14% of them found it 'Unfeasible'.

The arguments justifying the feasibility of this statement were that:

- much more culturally inclusive conversations about space exploration are needed to achieve more diversity of perspectives in the debates, perhaps by means of consultations at a national or regional level with a global outlook, as we have seen bad terrestrial examples of such exclusion;
- this is possible and only obstructed by the limitations of some special theory, e.g., the idea that only life has value, which may make Mars preservation difficult to justify in the public arena;
- this kind of dialogue is where the best potential for action can be generated;
- this is outreach, should policy and guidelines for preservation be developed.

The arguments justifying the unfeasibility of this statement were that:

- if all constituencies are included, mutually exclusive positions will be found, which makes agreement challenging;
- representatives of various nations should and will make these decisions, but it will not be with universal public support or even assent;
- there is limited attention to and support by the general public for these topics.

Refraining from judgment was justified by the following view:

- ideologies and cultural values are diverse, so it is unlikely that a sufficient granulation of views would be included in any supposedly broad consultation at a global level.

Regarding 'Statement 16: COSPAR's Panel on Exploration, advised by the Panel on Planetary Protection, undertakes oversight of a preservation framework for Mars', the main findings are as follows. The majority judgment was 'Definitely Feasible' (43%); a 'No Judgment' position was expressed by 21% of the respondents, while 21% of them found this statement 'Feasible', and 14% of them found it 'Unfeasible'.

The arguments justifying the feasibility of this statement were that:

- this could be done as an input to a broad and global dialogue among governments, private industry, and diverse public groups regarding preservation plans for Mars and to a collective international agreement on how to conduct Mars activities within such plans;
- this may likely happen, although COSPAR, as an NGO with a specialist audience, is more likely to advise governments on the implementation of such a framework;
- the COSPAR approach has proven successful in the past, so it would be ideal for COSPAR, with its three expert panels (Panel on Exploration, Panel on Potentially Environmentally Detrimental Activities in Space, Panel on Planetary Protection), to oversee a stronger protection regime for Mars;
- this is perhaps appropriate, but definitions of harm should be broadened beyond the narrow life-centric view assumed by COSPAR's planetary protection policy.

The arguments justifying the unfeasibility of this statement were that:

- COSPAR has a lot of history that might be problematic for a new authoritative oversight body, as it emerged during an era of US hegemony in space which is now coming to an end;
- many countries will not subject their sovereign authority to COSPAR's Panel on Exploration for oversight.

Refraining from judgment was justified by the following views:

- 'preservation' is a loaded term which, for many people, means no change and no use, usually facing considerable resistance, so 'conservation' is more appropriate and would be feasible;
- COSPAR will have no powers and will be ineffectual.

Regarding 'Statement 17: An Interplanetary Space Authority is created as a specialized UN agency responsible for governing a preservation framework for Mars', the main findings are as follows. The majority judgment was 'Feasible' (43%); a 'No Judgment' position was expressed by 29% of the respondents, while 14% of them found this statement 'Definitely Feasible', and 14% of them found it 'Definitely Unfeasible'.

The arguments justifying the feasibility of this statement were that:

- perhaps COPUOS or UNOOSA already is or does this;
- a specialized UN Agency would be needed for preservation to mirror COSPAR's role for science and it could also have responsibility for other celestial bodies;
- this may be a more powerful outcome from engaging thought leaders, community leaders, and experts first to develop a dialogue around long-term thinking for a variety of issues, including for space exploration and Mars settlement.

The arguments justifying the unfeasibility of this statement were that:

- although UNOOSA is such an agency, it has no powers and is ineffectual, because the five permanent members of the UN Security Council often interfere and veto any effective resolution or activity;

- many countries will not subject their sovereign authority to a specialized UN agency for governance, as exemplified by the particular absence of support from the United States in many activities developed through majority vote or consensus by and through the various UN agencies.

Refraining from judgment was justified by the following views:

- given the history of UN shortcomings, there is much to do first, including LEO, GEO, and the Moon;
- it may be better to emulate the Antarctic Treaty rather than the Outer Space Treaty;
- COSPAR works, so the invention of a new authority may be compelled by a political reason, but commercial and private actors must be integrated too;
- it all depends on the purview of this dedicated UN agency, on its authority, on its power and resources for serious oversight beyond the mere official approval of commercial proposals, and on whether it operates in ways more akin to the benefits-sharing model of the Moon Agreement, or the International Seabed Authority, or the Antarctic Treaty System or something other.

*4.3 Round 3 results*

The negligible attrition from Round 2 did not have a great effect on the good heterogeneity of the participants' group composition in Round 3. The attrition had little effect on the representation of various career stages (mean years of experience in the space sector: 25.1; standard deviation: 14.2). The synthesis of the group's collective professional expertise, as previously reported by the participants, was minimally affected thanks to existing overlapping, as presented in **Table 1**. Demographic diversity was also slightly affected: 15% of the participants were 30-39 years old, 23% of them were 50-59 years old, 38% of them were 60-69 years old, and 23% of them were 70 years old and above. Lastly, 77% of the participants had reported their sex as 'male', 15% of them as 'female', and 8% as 'other'.

This round was essentially an iteration of Round 2. Some of the Round 2 statements were reformulated and refined, based on the participants' short justifications, in order to deepen the discussion; these were marked as 'Reformulated Statements'. The 17 statements that were judged by the participants were again grouped into the three aforementioned sections and produced nominal categorical data whose relative frequencies are presented hereafter. Moreover, the open-ended form comments or critiques that were provided by the participants regarding the aggregate ratings, justifications, or content of each statement generated textual data that was analyzed using the initial coding steps of thematic analysis; the distilled coded arguments extracted from this data are presented hereafter for each statement.

Regarding 'Reformulated Statement 1: Private missions to Mars that are totally independent of government-sponsored missions will be an important driver of space activities in the future', the main findings are as follows. The majority position was 'Disagree' (31%); a 'No Judgment' position was expressed by 23% of the respondents; 23% of them stated that they 'Agree', 15% of them stated that they 'Strongly Agree', and 8% of them stated that they 'Strongly Disagree' with this statement.

The arguments justifying agreement with this statement were that:

- yes, in 100 years to 1,000 years from now, but not in less than 100 years;
- near term space activity and mission planning by actors like SpaceX, Blue Origin, Virgin, and others has already been and important driver, and it looks like it will continue to grow rapidly;
- they will become an important driver eventually, but in the short term state-sponsored exploration will continue to dominate;

- not 'totally independent' in the near term, and maybe not even in the long term, as it is expected that governments will be increasing the technology base on which private missions can build, creating a level of dependence, despite the fact that private actors may package space technology differently than government programs or conduct Mars missions unrelated to missions funded by government agencies.

The arguments justifying disagreement with this statement were that:

- it would be unacceptable to allow totally independent, unsupervised private movement of populations in space, because history has shown that such cases are connected with exclusive, oppressive, and undemocratic practices;
- political leaning and perhaps even nationality may influence judgment on this position, as some cultures, e.g. the US, are tightly connected to private enterprise while others not;
- although continued private efforts are usually helpful in developing new technology in a business context, involving crew or travelers is different, and the timeframe in the future is a crucial factor;
- one or two such missions might happen, but it is not yet clear what kind of activities would be sustainable for the private sector beyond cis-lunar space;
- it will be quite a while before there will be totally independent missions, as exemplified by the continuing government support for most airlines globally.

Refraining from judgment was justified by the following views:

- it depends on the timeframe, as it is definitely not possible in the near future, but it is conceivable in the far future, in a way similar to tourist companies sending trips to what was previously uncharted territories;
- it depends on the politics, as governments may decide to allow development of private deep-space capability, e.g., deep-space communications, or to keep such capability governmental as a means of keeping private companies 'in check';
- there are ethical, political, and funding issues with such missions, so, although conceivable, they are unlikely in the near term;
- such missions are probably inevitable, but the view that whoever gets to Mars first will make the rules, which is sometimes expressed by private actors, is dangerous.

Regarding 'Reformulated Statement 2: The progress of space activities on Mars will inflict significant harms on the Martian environment in the future', the main findings are as follows. The two majority positions were 'Agree' (31%) and 'Disagree' (31%); 23% of the respondents stated that they 'Strongly Agree' with this statement, while a 'No Judgment' position was expressed by 15% of the respondents.

The arguments justifying agreement with this statement were that:

- it is unavoidable under the status quo, although there is a question about the concepts and types of 'harm', 'damage', and so on, which are expected to broaden beyond 'life is all that matters' in the long run;
- harms will be suffered, but the question is to what degree, as it is hard to assess when a given harm becomes too significant;
- harm also depends on its duration;
- in response to all the justifications given for refraining from judgment in the previous round, it should be noted that, as exemplified by the nuclear and other industries, harm is inflicted regardless of precaution in the case of unforeseeable circumstances and because profits are considered more important than safety against harm which is seen as acceptable;
- the long-term consequences of actions are ignored;

- there are currently no regulations by any government (except Luxembourg and, perhaps, Belgium) that require space ventures to consider their adverse impacts on the martian environment except for biological contamination issues, while the US has policy in place to limit regulation of the growing private industry.

The arguments justifying disagreement with this statement were that:

- currently and for the foreseeable future, human activities on Mars will be insignificant compared to the martian environment, unless terraforming is attempted, so irreversible damage to natural environmental cycles on Mars or widespread destruction or devastation of unique landscapes or geological systems is unlikely, while finding life also seems unlikely and there doesn't appear to be any habitable places (aquifers) that can be contaminated;
- regulations and guidelines for responsible exploration will continue to guide the progress of space activities, although the length and width of their adherence to those guidelines will determine how far into the future significant harmful effects on the martian environment could occur;
- as responsible exploration continues to increase significantly our knowledge of Mars, more appropriate and cost-effective measures to protect the planet will be developed;
- it is difficult to confidently qualify human effects on the martian environment as truly harmful or beneficial in the long term, as, although human activities like resource use are very likely to cause environmental issues, perhaps some even unforeseeable, maybe they will eventually be a benefit to a far future martian biosphere.

Refraining from judgment was justified by the following views:

- it depends on the implementation of space activities on Mars;
- it depends on the definition of 'significant harms', because there will be impacts but their significance and harmfulness is uncertain.

Regarding 'Reformulated Statement 3: Space activities on Mars will reduce geopolitical conflict', the main findings are as follows. The majority position was 'Disagree' (31%); a 'No Judgment' position was expressed by 23% of the respondents; 23% of them stated that they 'Agree', and 23% of them stated that they 'Strongly Disagree' with this statement.

The arguments justifying agreement with this statement were that:

- there is a tendency for shared progress in space to reduce or mitigate conflict between the sharing partners;
- currently, space exploration is too expensive for individual nation states to tackle on their own, whereas there is not much to fight about in space right now;
- while initially nationalistic motives and profit will drive competition, the quest for Mars as another habitable planetary body has the potential of serving to promote more collaboration between nations on Earth and on Mars.

The arguments justifying disagreement with this statement were that:

- if Mars activities involve the reallocation of scarce resources, both tangible and intangible, then politics will be involved;
- in support of all the justifications for disagreement given in the previous round, the Outer Space Treaty has strong statements about cooperation, which has not yet reduced geopolitical conflicts on Earth (see the exclusion of China from US activities in space);
- as explained by justifications for disagreement given in the previous round, space activities are a minor aspect of international conflict;

- economic, environmental, and military competition will continue in this new commons in ways similar to how conflicts are generated on Earth;
- space on Earth continues to be dominated by defense interests, and, although the connection between settling Mars and national security is not directly obvious, a new arena for competition is not a great idea.

Refraining from judgment was justified by the following view:

- it is possible, but contingent upon how things unfold.

Regarding 'Statement 4: Preservation plans for Mars will enhance international space governance', the main findings are as follows. The majority position was 'Agree' (54%), followed by 'Strongly Agree' (31%); a 'No Judgment' position was expressed by 8% of the respondents, while 8% of them stated that they 'Strongly Disagree' with this statement.

The arguments justifying agreement with this statement were that:

- the timeline for exploration of Mars, especially preparing for a human future on Mars, will necessitate agreements between nations, and preservation planning for Mars provides a framework for enhancing such international cooperation and governance, at least to some extent, judging by the model of COSPAR's planetary protection policy;
- there is no real way to determine what sector(s) should predominate the relevant international discussions for decision-making, if humans indeed gain access to Mars (legal, scientific, for-profit, transportation, cultural, environmental, etc.);
- it is not obvious that the creation of sites for preservation on Mars will allow access for only a very small approved minority of actors, which will be undemocratic and unfair, as there is no specification of an actual driver to reasonably assume this;
- a democratically espoused plan for Mars needs preservation planning to safeguard the value in the space endeavor and to exclude only poorly-motivated and poorly-supported plans for overhasty human activities on Mars, including plans like settling Mars;
- reaching international consensus on the values to be preserved in the martian environment is a relatively low-stake way to test international agreements and contribute to good governance, because Mars is part of most people's cosmologies and worldviews.

The argument justifying disagreement with this statement was that:

- in support of all the justifications for disagreement given in the previous round, as recently exemplified by the Artemis Accords which are supported by a minority of countries that is not even significant, it is not really 'international' governance, but a domination of plans for Mars by the interests of the US, which also wields a UN veto, while around 200 countries are left out of access to space.

Refraining from judgment was justified by the following view:

- there may be zones on Mars that are provided varying levels of protection, one level of which may be 'preservation', i.e., no change in the environment, but Mars will not be 'preserved'.

Regarding 'Statement 5: Space activities on Mars with the goal of actually searching for life should progress faster', the main findings are as follows. The majority position was 'Strongly Agree' (38%); a 'No Judgment' position was expressed by 23% of the respondents; 15% of them stated that they 'Agree', while 15% of them stated that they 'Disagree', and 8% of them stated that they 'Strongly Disagree' with this statement.

The arguments justifying agreement with this statement were that:

- this is important to happen before human arrival at Mars, by analogy to the archeologists exploring a site before the construction team builds on the site;
- this comes down to a question of both life science and values, since humanity's valuing of life might prove very important in reframing humanity's understanding of its own place in the order of things, especially in the current period of environmental crisis;
- there is great wealth of scientific knowledge and understanding that can be gained regarding the origin and extent of life in the universe by exploring Mars, which may be both instrumentally and intrinsically valuable, especially if it represents a second genesis, and the means to accomplish this exist;
- 'faster' is not the right word, but this and other Mars science activities should be scaled up, sooner rather than later;
- the goal of searching for life need not be the sole objective of any mission or activity, but should be part of a set of objectives.

The arguments justifying disagreement with this statement were that:

- there is no need to rush a search for something that may not exist;
- getting the Earth more secure biologically is a greater priority;
- understanding the entire planet is a greater priority, because it's extremely difficult to completely sterilize spacecraft, and the risks of contamination are strong;
- searches should be reasonable, scientifically based, consistent with the Outer Space Treaty, and should occur in appropriate consultation with the other stakeholders and international decision-making groups.

Refraining from judgment was justified by the following views:

- the answer to the question of life might have scientific significance, but gaining that knowledge is not significant for a majority of the world;
- it's hard to rationalize the search for life on Mars;
- there is no overwhelming need to find extraterrestrial life;
- the rate should be slower and cautious to allow for new scientific insights to be derived and to be considered for future missions, as there will be no going back from contaminating potential life on Mars.

Regarding 'Statement 6: Research and technology development activities to enable crewed missions to Mars should progress faster', the main findings are as follows. The majority position was 'Strongly Agree' (31%); a 'No Judgment' position was expressed by 23% of the respondents; 23% of them stated that they 'Disagree', while 15% of them stated that they 'Agree', and 8% of them stated that they 'Strongly Disagree' with this statement.

The arguments justifying agreement with this statement were that:

- progress in R&D for crewed missions should be faster, because the discovery of extant life (or not) that represents a second genesis is a major driver on how we see the universe;
- the technologies to get an in situ presence on the martian poles should be sped up, because Mars polar science may be very important, for reasons that have nothing directly to do with large-scale settlement;
- the costs for human missions are small in relation to the key possibilities for Mars and the possible future paths they might define;
- critical knowledge gaps have been identified, a roadmap has been composed, and responsibly adding crews to the exploration of Mars will accelerate the pace and effectiveness of research;
- this should happen, but while taking into account limitations in our knowledge and prioritizing human safety and global cooperation.

The arguments justifying disagreement with this statement were that:

- there is no overwhelming need to enable crewed missions to Mars and it is not a priority;
- we are far from the point of diminishing scientific returns from uncrewed in situ and remote studies of Mars, and we need more information anyway for humans to safely go to Mars;
- a reasonable rate of exploration is warranted, but in context with sustainability of life on Earth;
- at present, crewed missions are a luxury that have not proven to be beneficial, but may be psychologically driven by a desire to ignore Earth's problems that seem difficult to solve.

Refraining from judgment was justified by the following view:

- regarding all the justifications for disagreement and for refraining from judgment given in the previous round, an additional concern is whether the Mars-required technology is useful and can be applied to situations and circumstances on Earth, besides its contribution towards Mars colonization, which would justify the expenditure as acceptable in the short term because of its potential social impact (e.g., medical technologies, recycling and sustainability, climate change solutions).

Regarding 'Reformulated Statement 7: National governments should have stronger political motivation and will for peaceful space activities on Mars', the main findings are as follows. The majority position was 'Disagree' (31%); a 'No Judgment' position was expressed by 23% of the respondents; 23% of them stated that they 'Agree', while 15% of them stated that they 'Strongly Agree' with this statement. It's worth noting that 8% of the respondents refrained from stating an answer to this statement.

The arguments justifying agreement with this statement were that:

- as long as policy is appropriately shaped, greater national government motivation and will would not be a bad thing and, if wisely applied, it could lead to technological and scientific advances that expand our knowledge of our planetary neighborhood and benefit life on Earth;
- governance has to be very pragmatic and fairly short-term, but including a recognition of our multi-generational place within a larger set of processes might be a good thing within bounds;
- rather than 'peaceful', those activities would better be 'cooperative'.

The arguments justifying disagreement with this statement were that:

- national governments should not have political motivations for supporting activities like settlement or resource exploitation on Mars, but they should only support scientific exploration activities;
- in support of all the justifications for disagreement given in the previous round, nationalism is an abhorrent basis for any activity and militates against the notions of international governance in space, as witnessed by the constant geopolitical and national oppositions ('peaceful economic activities') in humanity's only forum for international governance, i.e., the UN;
- Mars is not a global or national priority;
- it is inappropriate for mere political motivation to spur on something so risky, especially given that it would divert attention away from the needs and sustainability of Earth's environments and peoples.

Refraining from answering was justified by the following view:

- to prevent domination of Mars by private enterprise, national governments must take an interest, which is not expected to be driven by a desire to win votes but rather by national prestige and competition, essentially resembling a Cold War territory.

Regarding 'Reformulated Statement 8: The involvement of the private industry in shaping the objectives of space activities on Mars should be increased', the main findings are as follows. The majority position was 'No Judgment' (31%); 23% of the respondents stated that they 'Agree', while 23% of them stated that they 'Disagree', and 23% of them stated that they 'Strongly Disagree' with this statement.

The arguments justifying agreement with this statement were that:

- private industry involvement also means involvement in preservation plans;
- increased private involvement will benefit the leveraging of resources and accelerate the pace of crewed mission planning.

The arguments justifying disagreement with this statement were that:

- at the moment, the levels of co-operation and integration of private industry and government are about right and should continue;
- the private sector does seem to have a disproportionate influence upon the public discourse, but actual process remains primarily shaped by states and agencies;
- it is difficult to see either ideological or commercial motives for such an involvement, besides motives provided by a national government;
- corporations are not constituents, so it is not clear how their greater involvement would improve the democratic process rather than harm it and lead to worse outcomes for citizens;
- there should be balance, because private industry is not accountable to the public and may conceal a discovery, for example a discovery of life, for financial gain;
- so far, competitive private enterprise prioritizes profit and ideology over science, humanity, biosphere, and sustainability concerns;
- the involvement of the private industry should not be directed and should be left to the industry itself, as long as they do not destroy opportunities for future generations.

Refraining from judgment was justified by the following views:

- this involvement of the private industry should be in cooperation with national governments and other interest groups;
- this input should be reasonable, following assorted criteria and including more perspectives than just economic and industrial ones.

Regarding 'Reformulated Statement 9: Humanity should establish a long-term presence on Mars with the goal of becoming multiplanetary', the main findings are as follows. The majority position was 'No Judgment' (38%); 23% of the respondents stated that they 'Strongly Agree', 15% of them stated that they 'Agree', 15% of them stated that they 'Strongly Disagree', and 8% of them stated that they 'Disagree' with this statement.

The arguments justifying agreement with this statement were that:

- it is the rational defense to 'long-term' threats to the Earth;
- it is a worthy, if not necessary, goal in the longer term;
- perhaps in a few centuries from now, as there is no need to accomplish multiplanetary life anytime soon;
- becoming multiplanetary is one of the key ways of ensuring humanity's long-term survival and will likely drive more solutions for the problems we have already created here on Earth.

The arguments justifying disagreement with this statement were that:

- exploration is one thing, but to suggest 'long-term presence' is counter to human adaptation;
- humanity could have a long-term presence on Mars, but not with the goal of becoming 'multiplanetary', as this idea is different from past and present activities of migration in that it is strongly linked to a moral imperative of colonial expansion in a utopian arena where humanity can start afresh.

Refraining from judgment was justified by the following views:

- as indicated in the justifications for disagreement given in the previous round, the multiplanetary narrative is misleading, as, unless sci-fi technologies for moving beyond the Solar System are developed, or off-world habitats are constructed, humans will have a heavy centering upon the Earth, despite the various outposts and presence elsewhere;
- if the timescale of this long-term future is very deep, then 'humanity' may be comprised by another species of humans;
- establishing long-term presence on Mars is irrelevant in the nearest future;
- it may not be possible for physical and biological reasons that need to be scientifically determined;
- disagreement with this utilization of 'should';
- none of the justifications given in the previous round are particular convincing.

Regarding 'Statement 10: Preservation plans for Mars should include more values than just the protection of astrobiological science from biological contamination', the main findings are as follows. The majority position was 'Strongly Agree' (85%); 8% of the respondents stated that they 'Agree' with this statement, while a 'No Judgment' position was expressed by 8% of the respondents.

The arguments justifying agreement with this statement were that:

- planetary protection is insufficient should there be competing national or commercial interests at Mars;
- current understanding of what should be protected stems from various 20th century discourses which offer only a limited insight into the range of things that humans are likely to value in the future;
- Mars should be seen not just as a resource for humans but as an entity in a solar system setting that has values in its own right;
- timeframes need to be considered;
- activities that will not contribute to long-term scientific preservation and options for future generations should not be allowed, as human missions and exploitation activities have one chance to be done 'right';
- the whole of human activities and interest in Mars is not science or astrobiology;
- more knowledge regarding what is representative and what is unique is needed, and the potential questions and needs of future generations must be anticipated in the present, perhaps by applying to Mars the existing well-developed systems for assessing the value of cultural and natural entities on Earth.

Regarding 'Reformulated Statement 11: Inclusive broad and global dialogue among governments, private industry, scientific communities, and diverse public groups regarding preservation plans for Mars should be initiated promptly', the main findings are as follows. The majority position was 'Agree' (54%), followed by 'Strongly Agree' (38%); a 'No Judgment' position was expressed by 8% of the respondents.

The arguments justifying agreement with this statement were that:

- all the justifications for agreement given in the previous round are well made;
- broad international consensus on preservation can only be achieved by the appropriate dialogue;
- irreparable environmental damage that could occur by crewed or robotic missions must be avoided;
- COSPAR can be used as the starting point;

- there is always ongoing dialogue, but it should be further promoted, especially in a multi-stakeholder approach that includes people whose activities need to be contained rather than encouraged;
- although the current focus in on the Moon, a consistent approach could also include Mars and every solar system body, as many of the same issues are applicable.

Regarding 'Statement 12: International governmental actors reach collective agreement on how to conduct Mars activities within a preservation framework for Mars', the main findings are as follows. The majority judgment was 'Feasible' (62%), followed by 'Definitely Feasible' (31%); 8% of the respondents found this statement 'Definitely Unfeasible'.

The arguments justifying the feasibility of this statement were that:

- this has been demonstrated in the past through other activities like the Antarctic Treaty, Aviation Law, the Law of the Sea, COSPAR, etc.;
- treaties are entirely feasible, given the right political circumstances, so an environmental framework for Mars could be such an instance;
- an appropriate international body can facilitate such a broad collective agreement among multiple government actors;
- preservation is only one strategy of environmental and cultural heritage management, and is a loaded term, so the overarching term 'environmental management' can be used in this context of deciding, for example, which places on Mars to be preserved pristine (such as a subsurface salt lake) and which places to be developed (such as an area with some precious resource needed for survival), by weighting competing claims to find solutions and achieve sustainable outcomes.

The arguments justifying the unfeasibility of this statement were that:

- a single agreement might appear unfeasible in the present geopolitical climate;
- dominant actors are likely to dominate the different political and economic ideologies and cultural values of the many nations on Earth, as shown in the Artemis Accords;
- 'preservation' is a loaded term which, for many people, means no change and no use, usually facing considerable resistance, so 'conservation' is more appropriate and would be feasible;
- unlike conservation, preservation has national sovereign appropriation connotations, which will be problematic because access to all rather than a few is necessary and democratic.

Regarding 'Statement 13: National governments work on national space legislation regarding a preservation framework for Mars', the main findings are as follows. The majority judgment was 'Feasible' (62%); 15% of the respondents found this statement 'Definitely Feasible', and 15% of them found it 'Unfeasible'; a 'No Judgment' position was expressed by 8% of the respondents.

The arguments justifying the feasibility of this statement were that:

- a treaty, convention, or charter could be the first step, with national governments of signatories obliged to establish relevant national legislation in line with the requirements (as the 'Convention Concerning the Protection of the World's Cultural and Natural Heritage' or the 'Ramsar Convention on Wetlands of International Importance Especially as Waterfowl Habitat' operate);
- each nation would want its own perspective included in any internationally agreed framework, so this could be a precursor to a collective agreement;
- especially national frameworks by launching states may be one of the pathways for larger treaties that appears possible;

- this kind of legislation could be applied to all government, public and private actors planning Mars missions, strengthening the application of the Outer Space Treaty to Mars;
- this could be feasible for some national governments, but others may not find meaningful such a framework, or could even perceive it as restrictive and thus resist it;
- 'preservation' is a loaded term which, for many people, means no change and no use, usually facing considerable resistance, so 'conservation' is more appropriate and would be feasible;
- COSPAR and planetary protection follow this approach and provide an example that it is feasible;
- international treaties and obligations in theory override national legislation, but instances of withdrawal from treaty obligations and of national legislation that undermines the Outer Space Treaty, as exemplified by the Artemis Accords, are problematic and directly contravene international consensus.

The arguments justifying the unfeasibility of this statement were that:

- while national governments can think about their objectives and plans, any major decisions and actions regarding other planetary bodies should be coordinated at the international level;
- preservation, to most, means that nothing will be adversely impacted, which does not appear feasible.

Regarding 'Statement 14: Non-state actors, including commercial actors and NGOs, become involved in the development of a preservation framework for Mars', the main findings are as follows. The majority judgment was 'Definitely Feasible' (46%), followed by 'Feasible' (31%); a 'No Judgment' position was expressed by 23% of the respondents.

The arguments justifying the feasibility of this statement were that:

- this kind of legislation could be applied to all government, public and private actors planning Mars missions, strengthening the application of the Outer Space Treaty to Mars;
- this falls under inclusion, and it also makes sense to try to contain hostility to regulation through such inclusion, and mitigating the extent to which rules are seen as impositions rather than the expression of a consensus;
- as definition of a preservation framework is like defining standards, in that it is a negotiated and political process, non-governmental actors will want to participate and influence the final result, so they need to be part of this effort from the beginning for this to succeed, although their interests should not override or outcompete the ones of other stakeholders;
- non-state actors already are getting involved with COSPAR and planetary protection, so this provides an example that it is feasible;
- NGOs bring huge amounts of expertise, and commercial actors need to feel that it is important for them to be perceived as doing the right thing by the public, which is another stakeholder.

Refraining from judgment was justified by the following views:

- insufficient knowledge of this matter on the respondents' part;
- all actors (commercial, NGOs, state, etc.) have their own very narrow agendas, some of which are diametrically opposed;
- non-state actors should be involved in discussions, but in coordination with international bodies such as the UN.

Regarding 'Reformulated Statement 15: Public support for a preservation framework for Mars is earned through a broad consultation on a global level with experts in Mars activities and non-experts, including indigenous peoples of Earth', the main findings are as follows. The majority judgment was 'Definitely Feasible' (46%); 23% of the respondents found this statement 'Feasible', 15% of them found it 'Unfeasible', and 8% of them found it 'Definitely Unfeasible'; a 'No Judgment' position was expressed by 8% of the respondents.

The arguments justifying the feasibility of this statement were that:

- it's a political process, so broad public input, as well as inclusion of social movement (including indigenous) communities is required for successful completion;
- regional or national consultation with the public could be done to contribute to a global report;
- there is no special problem when it comes to the inclusion of mutually exclusive positions within the discussion, as this is a common part of policy discussions;
- including indigenous peoples is not just about being nice, but it is a practical matter, because some of these peoples have actual experience of extreme environments, of seeing themselves as belonging to extreme environments (and not at war with them), and as belonging to more than the Earth, and they are good shapers for multi-generational space projects as well;
- COSPAR and planetary protection provide an example that this is feasible.

The arguments justifying the unfeasibility of this statement were that:

- agreement will be challenging, because, if all constituencies are included, mutually exclusive positions will be found, as differences of opinion based on widely varying cultures and traditions are not compatible even within a single mono-culture;
- representatives of various nations should and will make these decisions, but it will not be with universal public support or even assent;
- there is limited attention to and support by the general public for these topics, as the average citizen can barely contribute to implementable frameworks for preservation on Earth;
- the notion of indigenous peoples is a variable one and changes through history, so perhaps at some point this might become ethnic nationalism.

Regarding 'Statement 16: COSPAR's Panel on Exploration, advised by the Panel on Planetary Protection, undertakes oversight of a preservation framework for Mars', the main findings are as follows. The majority judgment was 'Definitely Feasible' (38%), followed by 'Feasible' (31%); a 'No Judgment' position was expressed by 23% of the respondents, while 8% of them found this statement 'Unfeasible'.

The arguments justifying the feasibility of this statement were that:

- the COSPAR approach with planetary protection has worked well over several decades, so this existing good collaboration should result in effective planning, though they would need to initiate a new process to incorporate the changing situation;
- this is perhaps appropriate, but definitions of harm should be broadened beyond the narrow life-centric view assumed by COSPAR's planetary protection policy;
- COSPAR has just formed a Panel on Social Sciences and Humanities which is a good means of broadening the purview beyond planetary protection;
- COSPAR has a lot of history that might be problematic;
- including a broader range of people in the dialogue is likely to lead to calls for inclusion at all levels, including the formation of regulatory bodies;
- using existing regulatory and science groups can foster discussions and provide balanced analysis, but it is unclear how to turn that into anything 'implementable' in real-world terms;

- COSPAR would first need restructuring to make real and impactful change by engaging or enacting global dialogue to broadly promote such a framework for global adoption, before it is granted further powers or authority.

The arguments justifying the unfeasibility of this statement were that:

- COSPAR has a lot of history that might be particularly problematic for a new authoritative oversight body, as it emerged during an era of US hegemony in space which is now coming to an end;
- many countries will not subject their sovereign authority to COSPAR's Panel on Exploration for oversight;
- COSPAR, as an NGO with a specialist audience, is more likely to advise governments on the implementation of such a framework;
- definitions of harm should be broadened beyond the narrow life-centric view assumed by COSPAR's planetary protection policy.

Regarding 'Statement 17: An Interplanetary Space Authority is created as a specialized UN agency responsible for governing a preservation framework for Mars', the main findings are as follows. The majority judgment was 'Feasible' (46%); a 'No Judgment' position was expressed by 31% of the respondents; 8% of the respondents found this statement 'Definitely Feasible', while 8% of them found it 'Unfeasible', and 8% of them found it 'Definitely Unfeasible'.

The arguments justifying the feasibility of this statement were that:

- perhaps COPUOS or UNOOSA already is or does this;
- a UN body might be more broadly acceptable than a COSPAR-based body;
- a COSPAR-like body is needed for other than scientific exploration;
- it will be feasible, if it includes private actors as well as governments;
- UNOOSA and COSPAR could play this role, but a convention could establish a new authority.

The arguments justifying the unfeasibility of this statement were that:

- although UNOOSA is such an agency, it has no powers and is ineffectual, because the five permanent members of the UN Security Council often interfere and veto any effective resolution or activity;
- many countries will not subject their sovereign authority to a specialized UN agency for governance, as exemplified by the particular absence of support from the United States in many activities developed through majority vote or consensus by and through the various UN agencies;
- the particular history of the US not participating in authorities and agreements with which it disagrees is very problematic, but, if that changed, then UNOOSA could become such an Authority;
- an alternative model might be the World Trade Organization, although here again the US is asking China to stick to the rules, but won't do so itself;
- the Antarctic Treaty is too short-term, and retains the sovereign claims of various nations on the Antarctic Territory, in opposition to the non-sovereign national appropriation clause of the Outer Space Treaty, making it an unsuitable mode for outer space governance;
- this is not even possible on Earth and will be very complex for actual authority and governance.

Refraining from judgment was justified by the following views:

- COSPAR works, so the invention of a new authority may be compelled by a political reason, but commercial and private actors must be integrated too;
- insufficient knowledge of this matter on the respondents' part.

*4.4 Overall results*

After Round 3, each statement was evaluated for stability and for the potential emergence of consensus or dissensus, as suggested by (Belton, et al., 2019). As the sample size was small and the closed-ended, Likert-type response scales generated nominal responses, for each statement a 3X3 contingency table was created via cross-tabulation of its Round 2 and Round 3 responses, after combining similar judgment options into one composite option, e.g. 'Strongly Agree' and 'Agree' into one aggregated option of agreement; the very few cases of no response were recoded as 'No Judgment', while the responses of the one participant who dropped out after Round 2 were excluded from this analysis. Using the contingency tables, stability for every statement was evaluated by means of both the McNemar-Bowker chi-square test (Rayens & Hahn, 2000; von der Gracht, 2012) and Cohen's kappa (Holey, et al., 2007; Weir, et al., 2006; von der Gracht, 2012) as complimentary measures, in light of the small number of observations; this approach was followed because the calculation of the McNemar-Bowker chi-square focuses on the elements around the diagonal of the contingency table and ignores the diagonal elements, while Cohen's kappa assesses whether the counts along the diagonal are significantly large. In specific cases where one of the two statistics could not be meaningfully calculated because of the limitations, e.g. because of a high prevalence of zero observations, the other one is reported. Since the null hypothesis of the McNemar-Bowker chi-square test was that responses have not changed significantly between rounds, a significant p-value in the test ($p < 0.05$) would suggest that responses changed between rounds, while a non-significant outcome ($p > 0.05$) would indicate that the responses remained stable between rounds (Younas, et al., 2021; Weir, et al., 2006). Cohen's kappa levels were assessed for stability following (Holey, et al., 2007): a value of 0 indicates agreement between rounds by chance, while a value of 1 indicates perfect agreement. Moreover, following (Belton, et al., 2019), the consensus criterion that was established a priori was that in Round 3 at least 75% of the participants should provide similar judgment options, e.g. 'Strongly Agree' and 'Agree'. The final findings with respect to stability between Round 2 and Round 3 and the emergence of consensus or dissensus after Round 3 are presented hereafter for each statement.

Regarding 'Reformulated Statement 1: Private missions to Mars that are totally independent of government-sponsored missions will be an important driver of space activities in the future', the main findings are as follows. Chi-square = 5.00 with $p > 0.05$; this indicates stability. Kappa = 0.409; this level indicates moderate stability. The consensus criterion was not met.

Regarding 'Reformulated Statement 2: The progress of space activities on Mars will inflict significant harms on the Martian environment in the future', the main findings are as follows. Chi-square = 5.00 with $p > 0.05$; this indicates stability. Kappa = 0.444; this level indicates moderate stability. The consensus criterion was not met.

Regarding 'Reformulated Statement 3: Space activities on Mars will reduce geopolitical conflict', the main findings are as follows. Chi-square = 3.00 with $p > 0.05$; this indicates stability. Kappa = 0.567; this level indicates moderate stability. The consensus criterion was not met.

Regarding 'Statement 4: Preservation plans for Mars will enhance international space governance', the main findings are as follows. Chi-square = 1.00 with $p > 0.05$; this indicates stability. Kappa was not calculated. The consensus criterion was met by the similar judgment options 'Strongly Agree' and 'Agree' (85%).

Regarding 'Statement 5: Space activities on Mars with the goal of actually searching for life should progress faster', the main findings are as follows. Chi-square was not calculated. Kappa = 1.000; this level indicates perfect stability. The consensus criterion was not met.

Regarding 'Statement 6: Research and technology development activities to enable crewed missions to Mars should progress faster', the main findings are as follows. Chi-square = 3.00 with $p > 0.05$; this indicates stability. Kappa = 0.420; this level indicates moderate stability. The consensus criterion was not met.

Regarding 'Reformulated Statement 7: National governments should have stronger political motivation and will for peaceful space activities on Mars', the main findings are as follows. Chi-square = 2.00 with p > 0.05; this indicates stability. Kappa = 0.425; this level indicates moderate stability. The consensus criterion was not met.

Regarding 'Reformulated Statement 8: The involvement of the private industry in shaping the objectives of space activities on Mars should be increased', the main findings are as follows. Chi-square = 3.80 with p > 0.05; this indicates stability. Kappa was not calculated. The consensus criterion was not met.

Regarding 'Reformulated Statement 9: Humanity should establish a long-term presence on Mars with the goal of becoming multiplanetary', the main findings are as follows. Chi-square = 2.00 with p > 0.05; this indicates stability. Kappa = 0.768; this level indicates substantial stability. The consensus criterion was not met.

Regarding 'Statement 10: Preservation plans for Mars should include more values than just the protection of astrobiological science from biological contamination', the main findings are as follows. Chi-square = 0.00 with p > 0.05; this indicates stability. Kappa was not calculated. The consensus criterion was met by the similar judgment options 'Strongly Agree' and 'Agree' (92%).

Regarding 'Reformulated Statement 11: Inclusive broad and global dialogue among governments, private industry, scientific communities, and diverse public groups regarding preservation plans for Mars should be initiated promptly', the main findings are as follows. Chi-square was not calculated. Kappa = 0.480; this level indicates moderate stability. The consensus criterion was met by the similar judgment options 'Strongly Agree' and 'Agree' (92%).

Regarding 'Statement 12: International governmental actors reach collective agreement on how to conduct Mars activities within a preservation framework for Mars', the main findings are as follows. Chi-square = 0.00 with p > 0.05; this indicates stability. Kappa was not calculated. The consensus criterion was met by the similar judgment options 'Definitely Feasible' and 'Feasible' (92%).

Regarding 'Statement 13: National governments work on national space legislation regarding a preservation framework for Mars', the main findings are as follows. Chi-square = 1.00 with p > 0.05; this indicates stability. Kappa was not calculated. The consensus criterion was met by the similar judgment options 'Definitely Feasible' and 'Feasible' (77%).

Regarding 'Statement 14: Non-state actors, including commercial actors and NGOs, become involved in the development of a preservation framework for Mars', the main findings are as follows. Chi-square was not calculated. Kappa = 0.381; this level indicates fair stability. The consensus criterion was met by the similar judgment options 'Definitely Feasible' and 'Feasible' (77%).

Regarding 'Reformulated Statement 15: Public support for a preservation framework for Mars is earned through a broad consultation on a global level with experts in Mars activities and non-experts, including indigenous peoples of Earth', the main findings are as follows. Chi-square = 2.00 with p > 0.05; this indicates stability. Kappa = 0.409; this level indicates moderate stability. The consensus criterion was not met.

Regarding 'Statement 16: COSPAR's Panel on Exploration, advised by the Panel on Planetary Protection, undertakes oversight of a preservation framework for Mars', the main findings are as follows. Chi-square = 1.00 with p > 0.05; this indicates stability. Kappa = 0.547; this level indicates moderate stability. The consensus criterion was not met.

Regarding 'Statement 17: An Interplanetary Space Authority is created as a specialized UN agency responsible for governing a preservation framework for Mars', the main findings are as follows. Chi-square = 0.00 with p > 0.05; this indicates stability. Kappa = 0.480; this level indicates moderate stability. The consensus criterion was not met.

To summarize, all statements appear to have reached satisfactory stability in their responses from Round 2 to Round 3. The consensus criterion after Round 3 was met in the following 6 statements: 4, 10, 11, 12, 13, and 14. The outcome of the other 11 statements (1, 2, 3, 5, 6, 7, 8, 9, 15, 16, and 17) was dissensus.

## 5. Discussion and Conclusion

As mentioned before, the 17 statements that were judged by the panel of experts were grouped into three sections: 'expectations' (statements 1 to 4), 'preferences' (statements 5 to 11), and 'policy perspectives' (statements 12 to 17). It can be observed that 1 out of 4 'expectations' statements, 2 out of 7 'preferences' statements, and 3 out of 6 'policy perspectives' statements have reached stable consensus.

Overall, as per the stable consensus, the panel's judgments indicate the following. The panel anticipates that preservation plans for Mars will enhance international space governance. In addition, the panel is in favor of the inclusion of more values than just the protection of astrobiological science from biological contamination in preservation plans for Mars. On a similar note, the panel is in favor of the prompt initiation of inclusive broad and global dialogue among governments, private industry, scientific communities, and diverse public groups regarding preservation plans for Mars. With respect to potential courses of action, the panel finds achievable the collective agreement of international governmental actors on how to conduct Mars activities within a preservation framework for Mars. The panel also finds achievable the development of national space legislation by national governments regarding such a framework, as well as the involvement of non-state actors, including commercial actors and NGOs, in the development of the aforementioned framework. In other words, it appears that the panel foresees international cooperation on matters related to preservation plans for Mars, which, in turn, will enhance the overall international space governance. It also appears that the panel foresees the inclusion of varying actors, both state and non-state ones, in the development of a preservation framework for Mars, especially on a national level. To this end, it appears that the panel endorses the inclusion of a diversity of values to be protected via the aforementioned framework, and suggests the advancement of broad global dialogue on this matter including academic, industrial, governmental, and public actors.

Regarding the 'expectations' statements, the panel disagrees on: whether private missions to Mars that are totally independent of government-sponsored missions will be an important driver of space activities in the future; whether the progress of space activities on Mars will inflict significant harms on the Martian environment in the future; and whether space activities on Mars will reduce geopolitical conflict. The stable dissensus in those statements may indicate the presence of contrasting or opposing beliefs and perceptions about the probable future. Specifically, there seem to be varying projected imaginings of how the future may unfold, based on different assumptions regarding the timeframes involved, the continuity of present tendencies, the applicability and relevance of historical and adjacent analogies, the motives and complex interactions of the various stakeholders, the potential for preventive normative measures to be established, as well as the conceptual and epistemic uncertainties at play

Regarding the 'preferences' statements, the panel disagrees on: whether space activities on Mars with the goal of actually searching for life should progress faster; whether research and technology development activities to enable crewed missions to Mars should progress faster; whether national governments should have stronger political motivation and will for peaceful space activities on Mars; whether the involvement of the private industry in shaping the objectives of space activities on Mars should be increased; and whether humanity should establish a long-term presence on Mars with the goal of becoming multiplanetary. The stable dissensus in those statements may indicate the presence of contradictory or incompatible normative beliefs and value orientations that extend into the preferred future. Specifically, there seem to be varying projected imaginings of how the future should unfold, based on different assumptions regarding what has value and the various values at play, the opportunity costs of prioritization, the benefits and risks at play, the balance of short-term and long-term interests and needs, as well as the balance of Earth-centric and space-centric interests and needs, the importance of legitimacy, the importance of responsible research, the importance of critical junctures that can set specific trajectories, the applicability of the 'ought implies can' formula, the motives and complex interactions of the various stakeholders, and the need for including diverse perspectives.

Regarding the 'policy perspectives' statements, the panel disagrees on: the feasibility of public support for a preservation framework for Mars being earned through a broad consultation on a global level with experts in Mars activities and non-experts, including indigenous peoples of Earth; the feasibility of COSPAR's Panel on Exploration, advised by the Panel on Planetary Protection, undertaking oversight of a preservation framework for Mars; and the feasibility of an Interplanetary Space Authority being created as a specialized UN agency responsible for governing a preservation framework for Mars. The stable dissensus in those statements may indicate the presence of divergent or disparate ideological inclinations and worldviews that influence the appraisal of suitable future courses of action. Specifically, there seem to be varying ex-ante evaluations of the viability of putative policy outlooks, based on different assumptions regarding the degree of effective public participation, the need for including underrepresented cultures such as indigenous peoples, the historical role and efficacy of existing institutions, the level of centralization of authority, the juxtaposition of preservation and conservation, and the agendas of nation states.

With respect to the aforementioned insights, it should be noted here that this study was not without limitations; although the authors strived to also include experts affiliated with private corporations in the panel, this was not attained. The panel's experts were affiliated with governmental agencies, academic institutions, private consultancies, and advisory services in a personal capacity. Also, because of the makeup of the initial sample frame, the participants' affiliated organizations were geographically located in the Western world, which is a further limitation of this work.

In conclusion, this preliminary work helped reveal and unpack the differing expert views and positions advocated with respect to the interplay between future space activities and future preservation plans on Mars. The results revealed the diversity of the main pro and con arguments for opposing positions in the cases of expert dissensus on specific aspects of this issue, which appear to be contingent on different assumptions related to the possible futures imagined by the experts. The rich pool of these arguments can provide a wealth of preliminary questions to be further explored in future, more targeted, policy-relevant research, for example in a subsequent empirical investigation of potential disciplinary differences in judgments and argumentation using more in-depth methods, such as fuzzy cognitive maps and focus groups. In parallel, the presence of expert consensus on specific aspects of this issue can be utilized as precursor input to inform relevant international discussions towards the formulation of proactive policies that will contribute to the environmental governance of future activities on Mars.

## Acknowledgements

The authors are deeply grateful to the panel of anonymous experts who participated voluntarily in this study and provided their valuable input. This research did not receive any specific grant from funding agencies in the public, commercial, or not-for-profit sectors.